\documentclass[english,aps,a4paper,prd,groupedaddress,nofootinbib,preprint,floatfix]{revtex4-1}

\RequirePackage[dvipsnames]{xcolor}
\usepackage[utf8]{inputenc}
\usepackage{amsmath}
\usepackage{amssymb}
\usepackage{graphicx}
\usepackage{caption}
\usepackage{subcaption}
\usepackage{float}
\usepackage{multirow}
\usepackage{url}
\usepackage{hyperref}
\usepackage{cleveref}

\DeclareMathOperator*{\argmin}{arg\,min}
\newcommand{\AddrLIP}{%
 LIP -- Laborat\'orio de Instrumenta\c{c}\~ao e F\'isica Experimental de Part\'iculas, Escola de Ciências,\\
 Departamento de F\'isica, Universidade do Minho, 
 4701-057 Braga, Portugal}
 
\newcommand{\AddrWurz}{%
    Institut f\"ur Theoretische Physik und Astrophysik,
    Uni W\"urzburg \\
        Campus Hubland Nord, Emil-Hilb-Weg 22,
D-97074 W\"urzburg, Germany
}

\begin{document}

\count\footins = 1000

\title{Exploring Parameter Spaces with Artificial Intelligence and Machine Learning Black-Box Optimisation Algorithms
}
\author{Fernando Abreu de Souza}\email{abreurocha@lip.pt}
\affiliation{\AddrLIP}
\author{Miguel Crispim Rom\~ao}\email{mcromao@lip.pt}
\affiliation{\AddrLIP}
\author{Nuno Filipe Castro}\email{nuno.castro@fisica.uminho.pt}
\affiliation{\AddrLIP}
\author{Mehraveh Nikjoo}\email{mehraveh.n@gmail.com}
\affiliation{\AddrLIP}
\author{Werner Porod}\email{porod@physik.uni-wuerzburg.de}
\affiliation{\AddrWurz}
\today

\begin{abstract}
 Constraining Beyond the Standard Model theories usually involves scanning highly multi-dimensional parameter spaces and check observable predictions against experimental bounds and theoretical constraints. Such task is often timely and computationally expensive, especially when the model is severely constrained and thus leading to very low random sampling efficiency. In this work we tackled this challenge using Artificial Intelligence and Machine Learning search algorithms used for Black-Box optimisation problems. Using the cMSSM and the pMSSM parameter spaces, we consider both the Higgs mass and the Dark Matter Relic Density constraints to study their sampling efficiency and parameter space coverage. We find our methodology to produce orders of magnitude improvement of sampling efficiency whilst reasonably covering the parameter space. The code of this work is available in \url{https://gitlab.com/lip_ml/blackboxbsm}.
\end{abstract}

\maketitle

\section{Introduction}

Although the Standard Model (SM) of particle physics is a hallmark of scientific achievement, it does not provide the complete picture of the fundamental degrees of freedom of the universe, leaving some phenomena unexplained. To tackle this, multiple Beyond Standard Model (BSM) theories have been proposed to address a number of questions, which the Standard Model (SM) has failed to provide meaningful answers to, whilst successfully replicating all the features contained in the SM which have been verified experimentally. On the other hand, experiments like those at the Large Hadron Collider (LHC) at CERN are pushing the boundaries of validity of many BSM theories, while not providing so far unambiguous evidence for new phenomena beyond the SM.

In order to study the phenomenology of these BSM theories, vast parameter spaces need to be scanned to assess the values of parameters which are still valid, \emph{i.e.} not in contradiction with experimental data. Such models can reach $\mathcal{O}(100)$ free parameters. However, in general, out of the virtually infinite number of possible versions of the BSM model which are represented by points scattered across the parameter space of the theory, only a tiny fraction of these points will yield predictions which are in agreement with experimental data. For instance, the Minimal Supersymmetric Standard Model (MSSM) contains 105 new free parameters, leaving a more classical examination of its parameter space rather costly and extremely time-consuming. This type of validation task can be strikingly difficult to execute, depending on the physics of the model, the number of parameters involved and the number of experimental constraints considered. This is the High-Energy Physics realisation of a challenge known in data science as the \textit{curse of dimensionality}, which, in this context, means that the efficiency of this exploratory analysis drops exponentially with the number of the dimensions of the parameter space.

In this regard, data-driven approaches have offered new opportunities for the investigation of high-dimensional complex problems. In recent years, Artificial Intelligence (AI) and Machine Learning (ML) have steadily become part of the tool-set of HEP researchers~\cite{Feickert:2021ajf}, as their algorithms provide paradigm shifting capabilities for data and computationally intensive tasks. One such task is the validation of BSM theories through constraining the associated parameter space. Such task has seen recent efforts and developments of the employment of AI/ML algorithms to mitigate the burden of such scans in an attempt to increase sampling efficiency. Recent attempts at tackling this problem have deployed techniques such as deep neural networks \cite{ren2019exploring, staub2019xbit} to try to guess if a new point is valid; Bayesian neural networks \cite{Kronheim:2020vct} to try to predict the observable value for a given parameter space point; active-learning methods \cite{caron2019constraining,Goodsell:2022beo} to find boundaries of valid subspaces; and generative models \cite{Hollingsworth:2021sii} used to re-sample from a collection of valid points. However, these efforts often require a large amount of data to be gathered previously for machine learning training -- which presumably are hard to come by -- before they can be used to suggest new points with high-efficiency, effectively not solving the sampling bottleneck. 

In this work, we offer a new perspective to the sampling of new consistent model points by re-framing the problem as a Black-box Optimisation Problem and bypassing the need for an initial set of sampled data. We show the efficiency of a dynamic optimisation approach to the survey of two MSSM realisations, the constrained MSSM (cMSSM) and the phenomenological MSSM (pMSSM), both displaying a large reduction of the initial MSSM free parameters, by constraining the respective parameter spaces as to provide a realistic Higgs mass. For each case, we will further increase the sampling difficulty by demanding a realistic Dark Matter Relic Density.

This work is organised as follows. In~\cref{sec:reframing-problem} we re-framed the sampling problem as a black-box optimisation problem, by introducing the notion of a cost function of physical observables (themselves dependent on the parameter) that needs to be minimised. The physics cases are introduced in~\cref{sec:physics-cases}, where we define the models and the observables which we will use to constrain the parameter space. Next, in~\cref{sec:methodology} we develop the methodology to be used for the scans, namely we introduce three AI/ML based search algorithms used for black-box optimisation and how they work, as well as discussing how the scan was designed. The results of the scans and a comparison between different samplers is then discussed in~\cref{sec:results}. Finally, in~\cref{sec:conclusions} we draw the conclusions of our study and highlight the benefits and the shortcomings of the presented methodology, providing new directions of future work.

\section{(Re)Framing the Problem\label{sec:reframing-problem}}

The customary approach to validate Beyond the Standard-Model extensions against constraints and bounds on observables is to randomly sample a point, $\theta$, from the parameter space, $\mathcal{P}$, which is then passed onto a computational routine, $\mathcal{R}$, that computes the relevant observables, $\mathcal{O}(\theta)$. The observables are then compared to experimental data, namely to check if they are within bounds (for example if the mass of an exotic new particle is above collider limits) or within uncertainties (for example if the mass of a Standard Model particle is within its uncertainties). If the point agrees with experimental data it is kept as a valid point, otherwise it is discarded. Depending on the difficulty of the problem at hand, i.e.\ how likely or not is for a random point to fit the constraints, this process can take long periods of time to collect enough valid points. On top of that, the random sampling is rather wasteful from the point of view of resources as the information of invalid points is simply discarded and not used to improve the sampling efficiency.

Previous works~\cite{caron2017bsm, ren2019exploring, staub2019xbit} attempted to reduce the scanning overhead by only passing to the computational routines points with a higher chance of passing the constraints. In order to achieve this, they trained Machine Learning models to either predict the values of the observables, $\mathcal{O}$ (using a regressor) or to predict if a point falls within experimental bounds (using a classifier). Using this methodology, they achieve a higher efficiency in the computational routine step, as only promising sampled points go through. In either case, this amounts to add a novel step in the workflow, which is the Machine Learning model between the sampling and the computational routine steps. Therefore, a possible difficulty with this approach is that the Machine Learning component might not have learned the phase space well enough to properly filter good points. Or, in other words, the efficiency of this filtering step is bounded by the amount of points sampled.

Another attempt~\cite{Hollingsworth:2021sii}, also using Machine Learning models, is to use generative deep learning to produce likely valid points. The authors trained Normalising Flow Networks on a collection of valid points in order to learn their distribution to sample more, novel points, from the same distribution. Although this approach differs from the above, as the Machine Learning component does not act as a filter, it faces similar obstacles as these models need vast amounts of data to be trained, for example the authors used $\mathcal{O}(10^6)$ valid points, which could be hard to collect in highly constrained scans.

\begin{figure}
    \centering
    \includegraphics[width=0.5\textwidth]{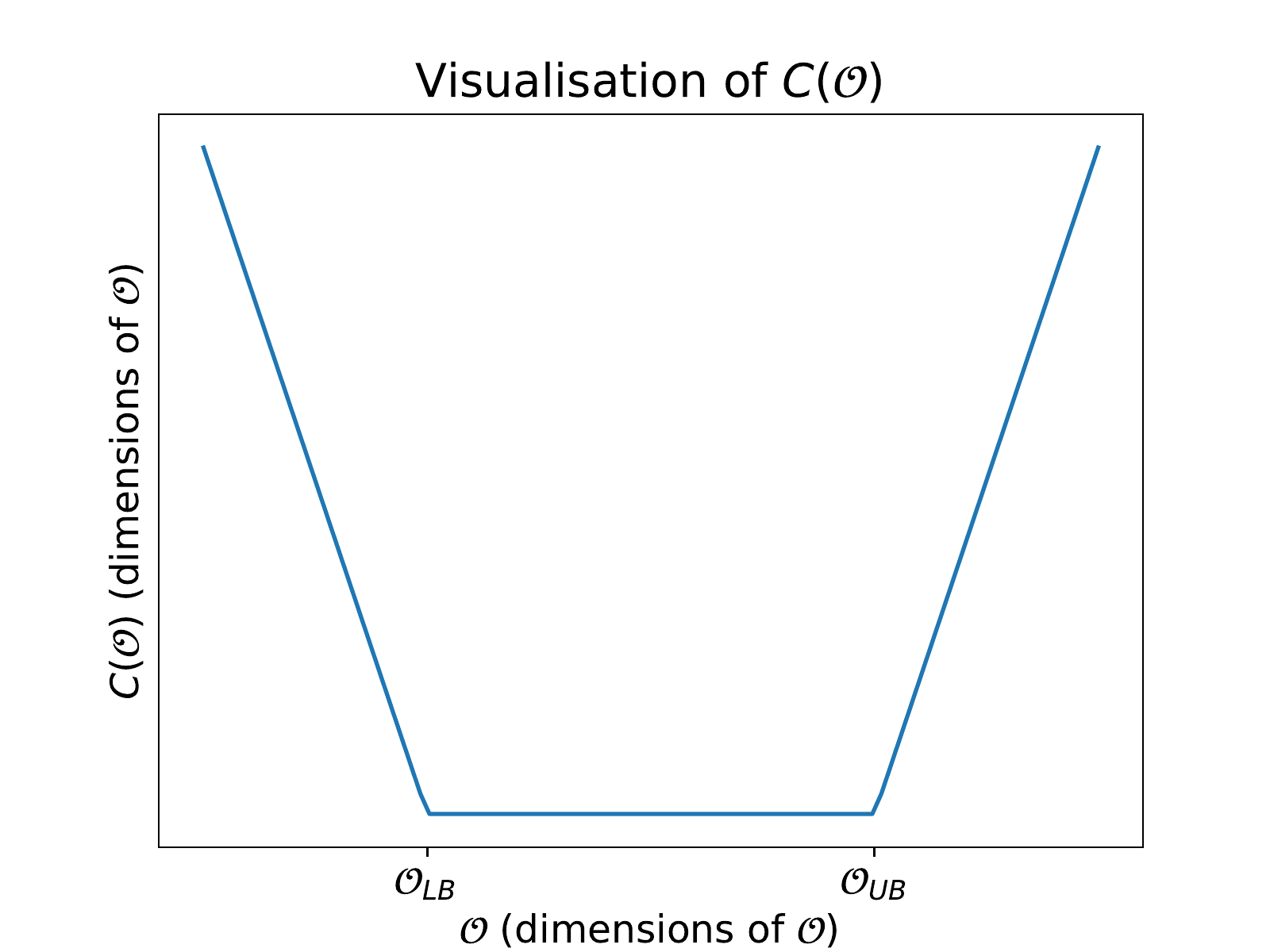}
    \caption{Shape of the constraint function for a single observable.}
    \label{fig:c_fun_plot}
\end{figure}

In this work we present a different approach by (re)framing the problem as a black-box optimisation problem to change the sampler itself. In order to shape the problem as an optimisation problem, we first notice that invalid points hold a wealth a information, namely the value of the constrained observables tells us \emph{how far} the point is from being valid. This can be captured by the constraint function, $C$:
\begin{equation}
\label{eq:c_func}
C(\mathcal{O}) = max(0, -\mathcal{O} + \mathcal{O}_{LB}, \mathcal{O} - \mathcal{O}_{UB}),
\end{equation}
where $\mathcal{O}_{LB}$ ($\mathcal{O}_{UB}$) is the lower (upper) bound of the observable $\mathcal{O}$. For example, if $\mathcal{O}$ is a Standard Model mass, say the Higgs mass, $\mathcal{O}_{LB/UB} = \mathcal{O}^{exp} \mp \sigma_\mathcal{O}$ with $\mathcal{O}^{exp} $ the observed central value of the mass and $\sigma_\mathcal{O}$\footnote{The notion of uncertainty depends greatly on the case-study. For example, one might want to include theoretical uncertainties, which do not have a statistical interpretation, or be more lenient and allow for up to 3 $\sigma$ deviations from each experimental bound.} the associated uncertainty. If, on the other hand, $\mathcal{O}$ is the mass of an exotic particle with experimental lower bound, $O^{exp}_{LB}$, then $\mathcal{O}_{LB} = O^{exp}_{LB}$ and $\mathcal{O}_{UB} = \infty$. However, we note that this function can be further expanded to included multi-dimensional exclusion regions, either from experiment or theory, with complicated shapes. To use those, one need to identify the \emph{inside} region where $C=0$ and the \emph{outside} region, where $C>0$ measures \emph{how far} the point is from the interior region. In~\cref{fig:c_fun_plot} we schematically show the shape of $C(\mathcal{O})$ for an observable with upper and lower bounds.

Considering now that the observables are functions of the parameters, $\mathcal{O}(\theta)$, and that the computational routines are in general black-boxes\footnote{There is some effort in the HEP community to produce end-to-end differentiable programming frameworks~\cite{baydin2020differentiable, heinrich2022differentiable, Kanwar:2020xzo} which would allow a purely differentiable treatment of the problem. However, for BSM model building, most of the available software exists either in non-differentiable frameworks or make use of non-differentiable routines.}, we have $C(\mathcal{O})=C(\mathcal{O}(\theta))=C(\theta)$. Therefore, finding valid points, $\theta^*$, can be defined in the usual way, as the valid points (i.e. that are inside the bounds), $\mathcal{V}$,
\begin{equation}\label{eq:old-scan-definition}
    \mathcal{V} = \{ \theta^* : \theta \in \mathcal{P} \text{ s.t. } C(\theta) = 0 \} \ ,
\end{equation}
which can be equivalently expressed as the minimisation statement
\begin{equation}\label{eq:new-scan-definition}
    \mathcal{V} = \{ \theta^* :  \theta \in \mathcal{P} \text{ s.t. } \theta^* = \argmin C(\theta) \} \ ,
\end{equation}
therefore, finding valid points for the constraints over $\theta$ amounts to minimise the function $C$ itself and so we can treat the problem as black-box optimisation problem. 

Multiple constraints can be combined using multiple $C$, one for each constraint. In principle, one could try to optimise against all constraints jointly as a \emph{multi-objective optimisation problem}, where one tries to find the so-called \emph{Pareto optimal} points\footnote{In practice this means that agreement with an observable cannot be improved without simultaneously worsening at least the agreement with another one.}. Here we will simplify this process and take the total constraint function as the sum of all individual constraints, as this new constraint function will still respect~\cref{eq:old-scan-definition} and~\cref{eq:new-scan-definition}, and allows us to use single-objective optimisation algorithms.\footnote{We performed an exploratory study on different prescriptions to join multiple constraints into a single function and could not observe any difference in early results. Further exploration of this choice might yield different results and is left as future work.}

In this work we will use the same constraints as in~\cite{Hollingsworth:2021sii}, namely the mass of the Higgs boson, $m_{h^0}$, and dark matter relic density, $\Omega_{DM} h^2$. The values of the upper and lower bounds can be seen in~\cref{tab:constraints}. Both observables are known precisely from the experimental side \cite{ParticleDataGroup:2020ssz}. However, their
uncertainty on the theory side is signifcantly larger amounting to about $\Delta m_h \simeq 3$~GeV \cite{Slavich:2020zjv}\footnote{Strictly speaking, the theory uncertainty on $m_h$is smaller within the CMSSM put for the sake of comparison we assumed that this uncertainty is model independent.}  and $\Delta \Omega h^2=0.2$ \cite{Boudjema:2011ig,Boudjema:2014gza,Harz:2016dql}.
We will aggregate both constraints by summing the individual constraint function for each constraint
\begin{equation}
C(m_{h}\cap \Omega_{DM}h^{2}) = C(m_{h^0}) + C(\Omega_{DM}h^{2}). 
\end{equation}
The resulting function will be the loss function,
\begin{equation}\label{eq:loss}
Loss(\theta)= C(m_{h^0}(\theta)) + C(\Omega_{DM}h^{2}(\theta)) \ ,
\end{equation}
which we will minimise using black-box optimisation algorithms presented in the next section.

\begin{table}
\centering
\begin{tabular}{lcc}
\hline\hline
Constraint             & $\mathcal{O}_{LB}$ & $\mathcal{O}_{UB}$ \\ \hline 
$m_h$            & $122$ GeV          & $128$ GeV          \\
$\Omega_{DM}h^2$ & $0.08$             & $0.14$ \\
\hline\hline            
\end{tabular}
\caption{Physical Constraints on the Higgs boson mass and Dark Matter Relic Density.}
\label{tab:constraints}
\end{table}

\subsection{Difference with Fits to Likelihoods\label{sec:likelihood}}

It is important to clarify the distinction between our approach and that of fitting the parameter space with likelihoods, see for example~\cite{Bernigaud:2021kpw}. When fitting the parameter space with likelihoods, one starts with  Bayes theorem
\begin{equation}
    p(\theta | \text{data}) \propto p(\text{data} | \theta) p(\theta) \ , 
\end{equation}
where $p(\theta | \text{data})$ is the posterior of the parameters (the probability of a choice of parameters, $\theta$, to be valid given the data), $p(\text{data} | \theta)$ the likelihood (a function which tells how likely it is the data given the choice of model and its parameters), $p(\theta)$ the parameter prior (which encodes prior distribution functions of the parameters), and we ignore the denominator which normalises the numerator. The fit is performed by making use of Monte-Carlo Markov Chain (MCMC) algorithms, which iteratively  adapt the sampling region, i.e. the prior, in order to find the posterior, i.e. to tell us how likely a certain choice of parameters is given the data.

In this approach, the likelihood functions can be constructed from some observational data, for example a Gaussian where the mean and the standard deviations are provided by an observation, or can be provided by the experiments themselves (see the efforts of some collaborations to provide likelihoods and other experimental data-derived statistical functions~\cite{Cranmer:2021urp}), over which the MCMC algorithm continuously samples and evaluates the priors in order to find the posterior distribution in a slow and computationally expensive process.\footnote{It is known that MCMC algorithms struggle with the so called \emph{curse of dimensionality}, i.e. with highly dimensional priors. There has been a considerable effort to mitigate this by using neural network approximators, which have been already used to perform these fits~\cite{Morrison:2022vqe}. The usage of neural networks has the added advantage of that they are differentiable and therefore easy to incorporate in MCMC algorithms that make use of derivatives, such as the Hamiltonian variation.} At the end of this process, a collection of points -- each retained according to its probability of being valid -- is obtained and from which a posterior distribution can be represented via histograms or other density estimators, with longer Markovian chains producing a better description of the posterior. MCMC fits discard many points, as they are only kept up to a probability of being valid, and can struggle to converge in high-dimensional spaces.

In our approach we are not looking for the posterior of the parameters given the data. This means that we are not concerned about how likely a point is given the data, i.e. the resulting distributions we obtain should not be interpreted as posteriors. We are concerned about how quickly and efficiently we can find regions and points of the parameter space which are valid. This means that we have to define what bounds on observables we are willing to accept, c.f.~\cref{tab:constraints}, and we do not have to concern ourselves with the explicit form of the likelihood. Indeed, the fact that we do not need a likelihood has its advantages, as our approach allows us to use bounds on masses or couplings of exotic physics by adding the appropriate constraint function, $C$, to better guide the sampler, whereas such information cannot be used in fits with likelihood functions.

\section{The physics models considered\label{sec:physics-cases}}

We take here the MSSM as underlying test model. On the one hand it remains an appealing SM extension, as it provides solutions to the most prominent shortcomings of the latter. In addition to solving the hierarchy problem related to the mass of the Higgs boson \cite{ATLAS:2012yve,CMS:2012qbp}, the model includes a viable candidate for the observed Cold Dark Matter (CDM) in the Universe, namely the lightest of the four neutralinos. 
On the other hand it can be formulated
either as a high scale model, where only
a few parameters are given, for example at the scale of grand unification. A prominent example is the constrained MSSM (cMSSM) \cite{Kane:1993td}, which is defined in terms of four parameters and the choice of a particular sign (phase). It can equally well be formulated as a low scale theory taking the soft SUSY breaking parameters
freely at the electroweak scale. A popular variant is the so-called pMSSM  \cite{MSSMWorkingGroup:1998fiq} which takes into account the most stringent constraints from low energy data by setting flavour mixing entries to zero and neglecting possible complex phases. 

Supersymmetric models are characterised via the
superpotential and the soft SUSY breaking Lagrangian. The superpotential of the MSSM is given as
\begin{align}
\label{eq:superpot}
W_{MSSM} &= -\varepsilon_{ab} \mu {\hat H}_1^a {\hat H}_2^b  
+\varepsilon_{ab}
\left(  {\hat H}_{1 }^a {\hat L}^{b} Y_e  {\hat E}^c +
 {\hat H}_1^a {\hat Q}^b Y_d  {\hat D}^c
+ {\hat H}_2^b {\hat Q}^a Y_u  {\hat U}^c \right)
\end{align}
$\varepsilon$ is the totally antisymmetric SU(2)
tensor, $Y_i$ are the Yukawa couplings and $\mu$
is the Higgs/higgsino mass parameter. The superfield $\hat F$ ($F=H_d,H_u,Q, L, D^c, U^c,E^c$) contains  the fermionic and bosonic degree of the field $F$. Here we have only
included terms conserving $R$-parity.
The soft Lagrangian is parameterised as
\begin{align}
	\begin{split}
		\mathcal{L}^{\rm MSSM}_{\rm soft} = 
		& - \frac{1}{2} \big( M_1\widetilde{B}\widetilde{B}+M_2\widetilde{W}\widetilde{W} 
		+ M_3\widetilde{g}\widetilde{g} + \rm{h.c.} \big) \\[1ex]
		& - m_Q^2\widetilde{Q}^{\dagger}\widetilde{Q} - m_L^2\widetilde{L}^{\dagger}\widetilde{L}
		- m_u^2\widetilde{U}^*\widetilde{U} - m_d^2\widetilde{D}^*\widetilde{D} 
		- m_e^2\widetilde{E}^*\widetilde{E}  \\[1ex]
		& - \big( T_U \widetilde{U}^*H_u\widetilde{Q} + T_D\widetilde{D}^*H_d\widetilde{Q}
		+ T_E \widetilde{E}^*H_d\widetilde{L}  \rm{h.c.} \big) \\[1ex]
		& - m_{H_u}^2 H_u^*H_u - m_{H_d}^2 H_d^*H_d - \big( b H_u H_d+{\rm h.c.} \big) \,.
	\end{split}
	\label{eqn:SUSY_breaking_Lagrangian}
\end{align}
where $\Tilde{\phi}$ denotes the superpartner of a generic SM particle $\phi$.
We neglect in the following all phases and flavour mixing entries. In this approximation
one can write the trilinear parameters $T_i$
as $T_i = A_i Y_i$. One has in total
even in this simplified version 
31 unknown parameters. Two of the four parameters in the Higgs sector ($\mu$, $b$,
$m_{H_u}^2$, $m_{H_d}^2$) are traded for
$M^2_Z$ and $\tan\beta=v_u/v_d$ where $v_{u,d}$ are the vacuum expectation values
of the Higgs bosons. In this way one ensures automatically that one complies with the precise measurement of the Z-boson mass and that one is in a minimum of the potential where SU(2)$_L \times$~U(1)$_Y$ is
correctly broken.\footnote{However, this does not necessarily imply that this is the global minimum of the potential, see e.g.~\cite{Camargo-Molina:2013sta,Camargo-Molina:2014pwa} and refs.~therein.}

In the following we will focus on the
Higgs mass and the dark matter relic density
as observables as already mentioned above. We summarise here a few key aspects of these observables as this will be helpful to understand some aspects of our findings. In contrast to the SM, the mass of the Higgs boson is not an independent quantity in supersymmetric models. Within the MSSM it is bounded from above
by M$_Z$ at tree level and large loop corrections are needed
to bring it to the observed value of about 125~GeV. The required large coupling is given by the top Yukawa coupling and consequently the largest contribution is given by loops containing top quarks or stops, see~\cite{Slavich:2020zjv} for a recent review. The relative large value of the Higgs mass m$_h$ implies that one needs either rather heavy stops and/or a large left-right mixing in the stop sector.
The mixing is controlled by the parameter $A_t$.
The observed relic density can be explained by
the lightest neutralino which is stable if it is the lightest supersymmetric particle (LSP) and if R-parity is conserved. Its dark matter properties depend strongly on its nature,
see e.g.~\cite{Roszkowski:2017nbc} for a recent review, which in turn depends on the hierarchy of the parameters $M_1$, $M_2$ and $\mu$. Besides its nature, which determines the annihilation rates into SM particles, the relic density will also depend on the nature of the next to lightest supersymmetric particle(s) as this might open co-annihilation channels if the mass difference is not too large \cite{Griest:1990kh}. Moreover, there is also the possibility of an s-channel resonance via the pseudoscalar Higgs boson if the mass of this Higgs boson is about twice the mass of the neutralino \cite{Griest:1990kh}.

We will use \texttt{SPheno} \cite{porod2003spheno,Porod:2011nf} for 
the calculation of the masses and mixing angles which serves as input for \texttt{micrOMEGAs} \cite{Belanger:2008sj,belanger2014micromegas}
which calculates the relic dark matter density.
The data transfer between these programs is handled using the SLHA format \cite{Skands:2003cj,Allanach:2008qq}. In \texttt{SPheno} the MSSM is matched onto the
SM at the scale $M_{SUSY}=\sqrt{m_{\tilde t_1}m_{\tilde t_2}}$ \cite{Staub:2017jnp} where $m_{\tilde t_i}$ are the masses of the two stops. In this way one ensures
a proper decoupling of the SUSY particles if their masses get very large compared to the electroweak scale.

\subsection{cMSSM}

The cMSSM is defined in terms of four parameters: at the scale of grand unification (GUT scale) one provides a common scalar mass parameter $m_0$ for the sfermions and Higgs bosons, a common trilinear coupling $A_0$ between sfermions and Higgs bosons as well as a common gaugino mass parameter $m_{1/2}$.
In addition one fixes $\tan\beta = v_u/v_d$ at the electroweak scale. The modulus of the superpotential parameter $\mu$ is fixed by the requirement of getting the correct value for M$_Z$ but its sign or more generally its phase is still a free parameter. We assume for this part of the investigation
$\mu >0$. We give in \cref{tab:cMSSM-paramaters} the ranges of the parameters considered as well
as the corresponding entry within the
SLHA format for the convenience of the reader.

The overall mass scale of the stops is roughly given by $\sqrt{m^2_0 + 4 m^2_{1/2}}$ and the left right mixing parameter $A_t \simeq - 2 m_{1/2} + 0.2 A_0$ in case of small $\tan\beta$. Approximate formulas for these parameters valid also for large $\tan\beta$ can be found in \cite{Blair:2002pg}. Thus, one
needs in general sizeable values of $m_0$
and  $m_{1/2}$ to explain the observed Higgs mass \cite{Slavich:2020zjv,Allanach:2004rh}.

\renewcommand{\arraystretch}{0.5}
\begin{table}
\centering
\begin{tabular}{llll}
\hline\hline
Parameter    & Values  & Description & \texttt{SPheno} input code  \\
\hline
$m_0$        & $[0, 10]$ TeV        &  Soft Scalar Mass & MINPAR: 1  \\
$m_{1/2}$    & $[0, 10]$ TeV        &  Soft Fermion Mass & MINPAR: 2\\
$A_0$        & $[ - 6 m_0, 6 m_0]$ & Trilinear Soft Coupling & MINPAR: 5\\
$\tan \beta$ & $[1.5, 50]$         & Tan Beta & EXTPAR: 25\\ \hline \hline
\end{tabular}
\caption{Parameters and their bounds of the pMSSM model. }
\label{tab:cMSSM-paramaters}
\end{table}

The required value of the DM relic density can only be achieved in particular slices of parameter space
where one has either co-annihilation or a Higgs-funnel resonance if the LSP is bino-like \cite{Jungman:1995df,Ellis:2012aa}. The co-annihilation usually requires a light stau or a light stop within the cMSSM \cite{Ellis:1999mm,Boehm:1999bj}. A wino-like LSP is not possible in this model but there is a slice where the LSP is higgsino like \cite{Jungman:1995df,Ellis:2012aa}.

\subsection{pMSSM}

In this model one defines the parameters
at the scale $M_{SUSY}$ neglecting
all CP phases and flavour mixing parameters. In addition one assumes that
the mass parameters of the first two generations sfermions are equal for particles with the same quantum numbers.
Moreover, the $A$-parameters of the first two generations are set to zero.
This amounts in 19 free parameters which
are summarised in \cref{tab:pMSSM-parameters} where we give again the corresponding entries for the SLHA convention in the last column.
The ranges for the parameters are chosen such that existing LHC bounds on the various supersymmetric particles are taken into account automatically. For certain combinations those bounds could be lowered but we do not expect that these additional points give additional features for the observables considered.

This additional freedom decouples completely 
the dependence of the two observables pMSSM
on the parameters. The stop mass
parameters are still the most important
ones for the Higgs mass. However, for the 
relic density several additional possibilities open up. Firstly, also the neutral wino becomes an accessible dark matter candidate. Secondly, in this class of models one can adjust the parameters such, that all electroweakly interacting supersymmetric partners can in principle be close in mass to allow for co-annihilation.
This is even true for squarks because the required small mass difference leads to very soft jets at the LHC which drastically reduces the bounds from direct searches \cite{LeCompte:2011fh,CMS:2019zmd,ATLAS:2020syg}. In particular light sleptons of the first generations can be light covering a part of the parameter space where the observed deviation of the anomalous magnetic moment of the muon can be explained \cite{Athron:2021iuf}.

\renewcommand{\arraystretch}{0.5}
\begin{table}
\centering
\begin{tabular}{llll}
\hline\hline
Parameter    & Values              & Description& \texttt{SPheno} input code \\
\hline
$|M_1|$        & $[0.05,  4]$ TeV &  Gaugino (Bino) mass& EXTPAR: 1       \\
$|M_2|$        & $[0.4, 4]$ TeV      & Gaugino (Wino) mass& EXTPAR: 2             \\
$M_3$        & $[1, 4]$ TeV        &  Gaugino (gluino) mass& EXTPAR: 3             \\
$|\mu|$        & $[0.4, 4]$ TeV      & Bilinear Higgs mass& EXTPAR: 23            \\
$|A_t|$        & $[0, 6]$ TeV       & Top trilinear coupling & EXTPAR: 11            \\
$|A_b|$        & $[0, 4]$ TeV       & Bottom trilinear coupling & EXTPAR: 12            \\
$|A_\tau|$     & $[0, 4]$ TeV       &  Tau trilinear coupling & EXTPAR: 13            \\
$m_A$        & $[0.1, 4]$ TeV      &  Pseudoscalar Higgs mass & EXTPAR: 26           \\
$\tan \beta$ & $[1,60]$            &   & EXTPAR: 25          \\
$m_{L_1}$    & $[0.1, 4]$ TeV      &  1st gen. l.h. slepton mass & EXTPAR: 31           \\
$m_{e_1}$    & $[0.1, 4]$ TeV      & 1st gen. r.h. slepton mass  & EXTPAR: 34          \\
$m_{L_2}$    & $m_{L_1}$           & 2nd gen. l.h. slepton mass& EXTPAR: 32            \\
$m_{e_2}$    & $m_{e_1}$           & 2nd gen. lr.h. slepton mass& EXTPAR: 35           \\
$m_{L_3}$    & $[0.1, 4]$ TeV      & 3rd gen. l.h. slepton mass & EXTPAR: 33            \\
$m_{e_3}$    & $[0.1, 4]$ TeV      & 3rd gen. r.h. slepton mass& EXTPAR: 36            \\
$m_{Q_1}$    & $[0.7, 4]$ TeV      & 1st gen. l.h. squark mass& EXTPAR: 41            \\
$m_{u_1}$    & $[0.7, 4]$ TeV      &   1st gen. r.h. u-type mass & EXTPAR: 44         \\
$m_{d_1}$    & $[0.7, 4]$ TeV      &  1st gen. r.h. d-type mass& EXTPAR: 47           \\
$m_{Q_2}$    & $m_{Q_1}$           &  2nd gen. l.h. squark mass & EXTPAR: 42          \\
$m_{u_2}$    & $m_{u_1}$           &  2nd gen. r.h. u-type mass & EXTPAR: 45          \\
$m_{d_2}$    & $m_{d_1}$           & 2nd gen. r.h. d-type mass & EXTPAR: 48            \\
$m_{Q_3}$    & $[0.7, 4]$ TeV      &  3rd gen. l.h. squark mass   &  EXTPAR: 43       \\
$m_{u_3}$    & $[0.7, 4]$ TeV      & 3rd gen. r.h. u-type mass & EXTPAR: 46        \\
$m_{d_3}$    & $[0.7, 4]$ TeV &  3rd gen. r.h. d-type mass & EXTPAR: 49          \\ \hline\hline
\end{tabular}
\caption{Parameters and their bounds of the pMSSM.}
\label{tab:pMSSM-parameters}
\end{table}

\section{Samplers and Methodology\label{sec:methodology}}

Having reframed the parameter space scan as an optimisation problem, and the physics cases that we will use in this work, we now present the samplers and the HEP computational routines that we will use.

The three sampling algorithms presented here, in addition to the random sampler that we will use as a baseline to compare their behaviour, operate in different ways and are representative of big classes of black-box optimisers. The purpose of using these three is to evaluate and assess how different approaches to black-box optimisation can impact the final result in terms of both sampling efficiency, i.e. how easily they produce valid points, and coverage of the parameter space, i.e. how much of the parameter space was explored and if the samplers are focusing on subsets of it. Indeed, these two characteristics present two opposing forces, which in Machine Learning and Artificial Intelligence literature is commonly known as \emph{exploration-exploitation trade-off}, where the former accounts for the capacity to explore the breadth of the parameter space, whereas the latter accounts for the inclination of an algorithm to exploit the information to get to a minimum (which could be local) as fast as possible.

As the approach presented herein is agnostic of the physics case being studied, and considers the HEP computational routine to be a black-box function, it is also important to point out that all algorithms used in this work are gradient-free, i.e. they do not rely on any gradient computation of the loss function. This is important as our loss function is a black-box function produced by the HEP routine which generally cannot be differentiated. In principle, one could compute numerical derivatives by evaluating in the infinitesimal neighbourhood of a point, however this would lead to too many black-box routine evaluations and to slower sampling speeds. Alternatively, one could produce a transparent box routine through which derivatives could be computed. Such approach, usually referred as \emph{differential programming}, would allow for different approaches making use of auto-differentiation such as those usually used in neural networks training. Unfortunately, this represents a change of paradigm in routine development, which is not yet customary in HEP and therefore outside the reach of this work.

\subsection{Tree-structured Parzen Estimator}

The Tree-structured Parzen Estimator (TPE)~\cite{TPE1, TPE2, TPE3} is a Bayesian optimisation algorithm. Such algorithms are composed of primarily two components: a surrogate model and an acquisition function. The surrogate model is a probabilistic model which iteratively approximates, i.e. learns, the cost function produced by the black-box, i.e. it approximates $p(Loss(\theta)|\theta)$. The acquisition function is a prescription to choose which point, as sampled using the information gathered by the surrogate model, is used to evaluate the black-box in the subsequent iteration.

Due to the probabilistic nature of the surrogate model, Bayesian optimisation algorithms have a natural predisposition to explore the parameter space early on, when few points have been sampled and the uncertainty about the cost function is high. As more points are used to learn the cost function, the acquisition function tends to prefer better points more confidently, moving the algorithm to an exploitation phase.

Each Bayesian optimisation algorithm has its own design for the surrogate model and acquisition function. The TPE uses Bayes theorem starting from the surrogate model
\begin{equation}
    p(Loss(\theta)|\theta) = \frac{p(\theta|Loss(\theta))p(Loss(\theta))}{p(\theta)}
\end{equation}
which is simplified by separating the points into two densities, one for good points, $l(\theta)$, and another for bad points, $g(\theta)$,
\begin{equation}
    p(\theta|Loss(\theta)) = \left\{ \begin{array}{cc} l(\theta), & \text{if} \quad Loss(\theta)<Loss^{*} \\ g(\theta), & \text{if} \quad Loss(\theta) \geq Loss^{*} \end{array}\right. \ ,
    \label{eq:proba}
\end{equation}
where $Loss^*$ is a cutoff value which splits points into good and bad.\footnote{Notice that in our case the black-box is deterministic, i.e. $p(Loss(\theta))$ is $1$ if the point has produced physical observables, and $0$ if it is not physical, i.e.~if \texttt{SPheno} does not produce a valid spectrum.} The distinction between good and bad is made through a quantitative heuristics built-in  routine, see~\cite{TPE1} for details,\footnote{The prescription to define $Loss^*$ is akin to a rolling quantile which becomes progressively smaller as the number of iterations grows.}  and the densities $g(\theta)$ and $l(\theta)$ are approximated using Gaussian Mixture Models. The crucial intuition is that sampling is performed on the good point distribution, and the quality of a new sampled point, $g(\theta^\prime)$, is a function of the likelihood ratio between both densities, $g(\theta^\prime)/l(\theta^\prime)$. Points which have a high likelihood ratio between both densities are kept, given to the black box, and the process repeats until a limit of trials has been performed. Early on, both distributions will be similar and diffuse, leading to a high exploration of the space. As more points allow for a better distinction between good and bad points, TPE will start to favour exploitation of the good points distribution. It is important to note that the value of the loss, $Loss(\theta)$, is only used to separate points using a heuristic cutoff value, i.e.~TPE does not learn $p(\theta|Loss(\theta))$ as it happens with other Bayesian optimisation algorithms. In other words, the value of the loss is only used to \emph{sort} the points, an operation which is independent of the nominal order of magnitude of the value of the loss function.

\subsection{Nondominated Sorting Genetic Algorithm II}

Nondominated Sorting Genetic Algorithm II (NSGA-II)~\cite{deb2002fast} is a genetic evolutionary algorithm. In genetic algorithms, the values of a parameter space point are encoded as genes, and a population is a collection of such points. Each population is evaluated by passing its members through the black-box, comprising a generation, and the members are ranked by the value of the respective loss. A new generation is produced by keeping the best elements, the \emph{elite}, and new elements are produced through \emph{offspring}, where genes are exchanged between two parents via \emph{cross-over} to produce a new member, exploiting the features of the elite parents. When new members are generated, \emph{mutations} can be applied to some genes (i.e. values of some of the parameters) randomly to increase exploration by applying Gaussian noise to the values of the parameters. As with any genetic algorithm, NSGA-II uses $Loss(\theta)$ to sort the members of the population to select the \emph{elite} that will produce the offspring.

In NSGA-II the members of the population are first sorted into groups regarding their loss function performance, and then further sorted by crowding distance to mitigate the risk of getting the population stuck in a local minima. NSGA-II is specially crafted for multi-objective optimisation problems. For single objective, as we perform here, it resembles a traditional genetic algorithm. The study of its performance and behaviour for multi-objective problems is left for a future work.

\subsection{The Covariance Matrix Adaptation Evolution Strategy}

The Covariance Matrix Adaptation Evolution Strategy (CMA-ES)~\cite{hansen2016cma} belongs to the class of population-based optimisation algorithm that do not implement genetic encoding to produce offspring. Instead, the algorithm samples new candidate points from a multivariate normal distribution, for which the mean -- that controls the direction of the evolution -- and the covariant matrix -- which captures the relations between parameters -- are adapted, i.e. learned, from the previous points. This is the sense where this is an evolutionary algorithm, as new points are produced through the information of the previous ones, but there is no direct parent to offspring genetic crossover, instead the new members of the population are derived from moving statistics.

The mean of the distribution is updated as to maximise the likelihood under the multivariate normal distribution of the best performing points. More specifically, the mean vector of the multivariate normal is updated through a rolling mean with the best points (usually half of the population). CMA-ES is expected to converge rapidly, as the (approximate) covariant matrix works as a proxy for the second derivative of the loss function, i.e. the Hessian, resembling a higher-order optimisation process. Intuitively, CMA-ES can be thought as of a herd of animals descending from the mountains, meeting in the valley, and moving together to the plane.

Although it uses a multivariate normal, CMA-ES is fundamentally different to TPE. In TPE a Gaussian Mixture Model is used to approximate point density, from which new points are sampled. Gaussian Mixture Models can fit multi-modal distributions, and provide a rich description of point density. On the other hand, a single multivariate normal, as used in CMA-ES, can only describe a single mode from which new points are then suggested. In particular, CMA-ES will focus on valid points around the current best mean, whereas TPE can maintain information of all previously tried points.

\subsection{Implementation details}

We have introduced three different black-box optimisation algorithms that cover three distinct classes: a Bayesian optimisation algorithm, a genetic algorithm as well as an evolutionary algorithm. This will allow us to explore the differences and nuances of each algorithm when applied to our problem. We now describe how our experiment was conducted.

For the numerical routines to compute physical observables, we have used \texttt{SPheno}-4.0.5~\cite{porod2003spheno} and \texttt{micrOMEGAs}\textunderscore 5.2.13~\cite{belanger2014micromegas}, in order to calculate the Higgs mass and dark matter relic density, respectively. We compute the mass spectrum using \texttt{SPheno} GUT scale input parameters for cMSSM (c.f.~\cref{tab:cMSSM-paramaters}), and SUSY scale for the pMSSM (c.f.~\cref{tab:pMSSM-parameters}). \texttt{SPheno} output spectrum files are used as inputs of \verb|micrOMEGAs| to calculate the dark matter relic density. We performed two parallel studies, with and without dark matter relic density constraint, while keeping the Higgs mass constraint for both of the studies.\footnote{The physics choice was made as to have a similar study to~\cite{Hollingsworth:2021sii}. However, their implementation relies on \texttt{SoftSUSY} version 4.1.0, whose routines to compute the parameters relevant to the Higgs mass differ, leading to lower sampling efficiencies. Nonetheless, we decided to keep these physics cases.}

The parameter spaces have been sampled and the loss optimised using \verb|Optuna|\textunderscore 2.8.0~\cite{optuna}, with the built-in Random, TPE, NSGA-II, and CMA-ES  samplers. We changed the default settings for the TPE sampler to \verb|multivariate=True|, in order to learn the correlations between the variables, and for the CMA-ES sampler to \verb|restart_strategy='ipop'|, which is a heuristic to restart the multivariate normal if convergence is seemingly stuck in a local minimum.

We did not sample directly from the parameter space definitions in~\cref{tab:cMSSM-paramaters} and~\cref{tab:pMSSM-parameters}. Instead, we sampled from a hyper-cube of size 1, which we call the box parameter space, $\mathcal{\hat P}$, which has the same dimension as the physical parameter space, $\mathcal{P}$. A box parameter space point, $\hat \theta \in \mathcal{\hat P}$, is then reshaped to be in $\mathcal{P}$ before being fed to the computational routine.\footnote{These transformations are mostly linear transformations to recentre and resize the interval from $[0,1]$ to the intended range. The exception being the parameters sampled from two disjoint intervals. Take for example the $\mu$ in the cMSSM case has values over $[-4, -0.4] \cup [0.4, 4]$ TeV. We first sample from $\hat\mu \sim [0.1]$, then reshape it to include negative numbers $\hat \mu = 2\times(\hat \mu - 0.5)$, then we keep its sign aside, and rescale and recentre its value to match the desired interval $\mu = \text{sign}( \mu^\prime)(|\mu^\prime| \times 3600+400)$. This way we avoid having to perform a separate sampling for the sign and all parameters are sampled from $[0,1]$.} This allows us to treat all the parameters as ranging the same nominal values, in this case between 0 and 1, to better derive comparing metrics, discussed bellow. We notice that the map is isomorphic, so a point $\hat \theta$ in $\mathcal{\hat P}$ maps to only one point in $\mathcal{P}$ and vice-versa, so they can be though as the same.

In early exploratory runs, we observed that the convergence speed for the TPE became progressively slower as the number of successive trials reached a few thousands. This is understood as the surrogate model in TPE, a Guassian Mixture Model, is known to have a high computational complexity, which makes it forbiddingly slow for long runs. In order to mitigate this, each scan for each sampler was limited to 2000 sequential steps, called trials, and repeated 500 time, which we call episodes, totalling one million points for each combination.

\subsection{Evaluating the Samplers}

In order to compare the samplers, we developed three different metrics. The first one, efficiency, is just the percentage of valid points found by the sampler
\begin{equation}
    \text{Efficiency} = \frac{\text{\# valid trials}}{\text{\# total trials}} \ .
\end{equation}
This is the most intuitive metric to compare samplers, as we want highly efficient samplers to tackle difficult constraints. However, we need to have a measurement on how the sampler is exploring the parameter space. We need a quantitative way of measuring how much of the parameter space each sampler has explored. To do this, we introduce two metrics.

The first metric to measure the width of the exploration is the mean euclidean distance between the sampled valid points. A sampler that explores narrow regions of the parameter space is expected to produce smaller mean distances between sampled valid points, whereas an exploration oriented sampler will produce high mean distances:
\begin{equation}
    \text{Mean Euclidean Distance} = \mathbb{E}_{\hat \theta_i, \hat \theta_j \in \mathcal{V}} \left[\sqrt{(\hat \theta_i-\hat \theta_j)^2}\right]\ ,
\end{equation}
where $\hat \theta_i$, $\hat \theta_j$ are any two points in the valid region of the parameter space, $\mathcal{V}$, as seen in the box parameter space, $\hat P$. The reason why this metric is obtained in the box parameter space is that higher nominal values would dominate the value of the distance, and dilute the impact of sparser distributions in smaller valued parameters. For a hyper-cube of dimension $d$ and size 1, the maximal distance between two points is given by the longest diagonal, $\sqrt{d}$, and it serves as gauge to the size of the box parameter space.

The second metric to measure the exploration is the Wasserstein distance (WD). Given two univariate distributions, $f(u)$ and $g(u)$ over the same domain, $u\in U$, and their cumulative distribution functions, $F(u)$ and $G(u)$, the Wasserstein distance between the two distributions is
\begin{equation}
    WD(f,g) = \int_U | F(u) - G(u)| du \ , 
\end{equation}
and measures how different the two distributions are. We will use this to measure how much of the parameter space is being covered by different samplers. To this effect, we compute WD for each parameter distribution of valid points against the uniform distribution, which the cumulative distribution function is just the straight line starting at the origin and ending at $(max(u), 1)$.\footnote{In fact, we computed over the distributions of the parameter values in the box space to simplify the process, where the endpoint is $(1,1)$.} Since the uniform distribution over a parameter is the maximal coverage possible in that dimension of the parameter space, this quantity measures how far off a distribution of valid points is from covering all possible values.

We notice however, as it was highlighted in~\cref{sec:likelihood}, our goal is not to fit the posterior distribution of the points, therefore this metric should not be taken as a dissimilarity measurement between the obtained distributions here and distributions obtained through a fit with likelihoods. Instead, this metric is a proxy to how far a sampler is from exploring the whole parameter space. We also note that the distributions of the random sampler are not expected to have vanishing WD with the uniform distribution, as the random sampler parameter distributions of valid points are distorted by the constraints and therefore will not be uniform distributions themselves.

The pairwise euclidean distances were computed using \verb|numpy|~\cite{van2011numpy} \verb|pdist| function. The cumulative distribution functions of the parameters were computed using \verb|statsmodels|~\cite{seabold2010statsmodels} \verb|ECDF| class. The Wasserstein distance was computed using \verb|SciPy|~\cite{virtanen2020scipy} \verb|wasserstein_distance| function. Data manipulation was done with \verb|pandas|~\cite{jeff_reback_2020_3715232}, and for data visualisation we used \verb|matplotlib|~\cite{hunter2007matplotlib}, \verb|seaborn|~\cite{waskom2021seaborn}, and \verb|mplhep|~\cite{andrzej_novak_2022_6332486}.

\section{Results\label{sec:results}}

We now present the results of the scans produced with the different samplers for the different Physics cases. For both the cMSSM and the pMSSM as introduced in~\cref{sec:physics-cases}, we performed two scans: one with the Higgs mass constraint only, and another with both the Higgs mass and the dark matter relic density constraints, with bounds defined in~\cref{sec:reframing-problem}.

\subsection{Target Observables and Sampled Parameters}

In this section we present the distributions of the target observables and scatter plots of some the parameters.

\subsubsection{cMSSM}

\begin{figure}
\centering
\begin{subfigure}{.4\textwidth}
  \centering
  \includegraphics[width=1.0\linewidth]{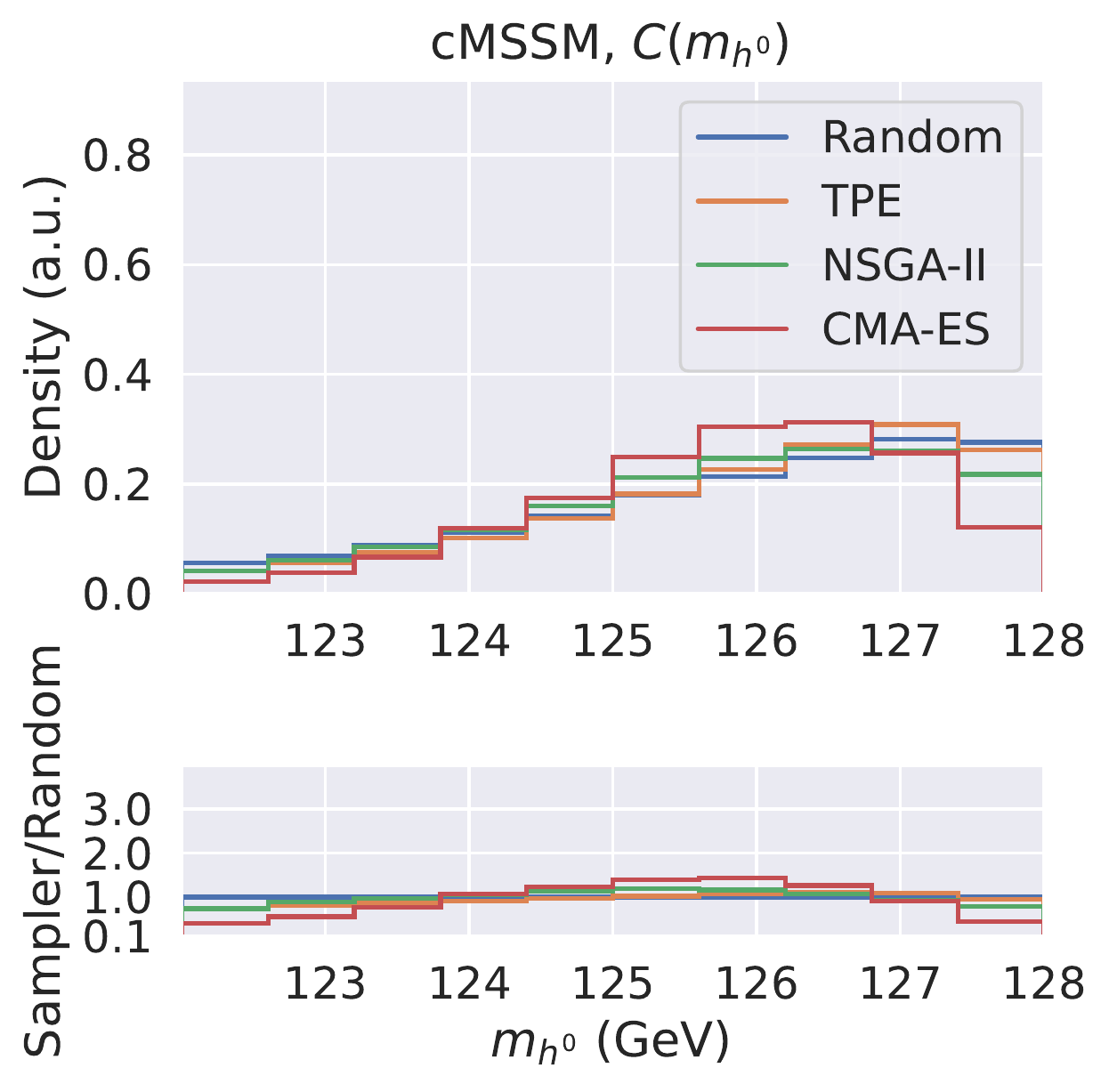}
   \caption{Higgs Mass distribution for the scan with the Higgs mass constraint.}
  \label{fig:cMSSM_mh0_wo_mo}
\end{subfigure}%
\begin{subfigure}{.4\textwidth}
  \centering
  \includegraphics[width=1.0\linewidth]{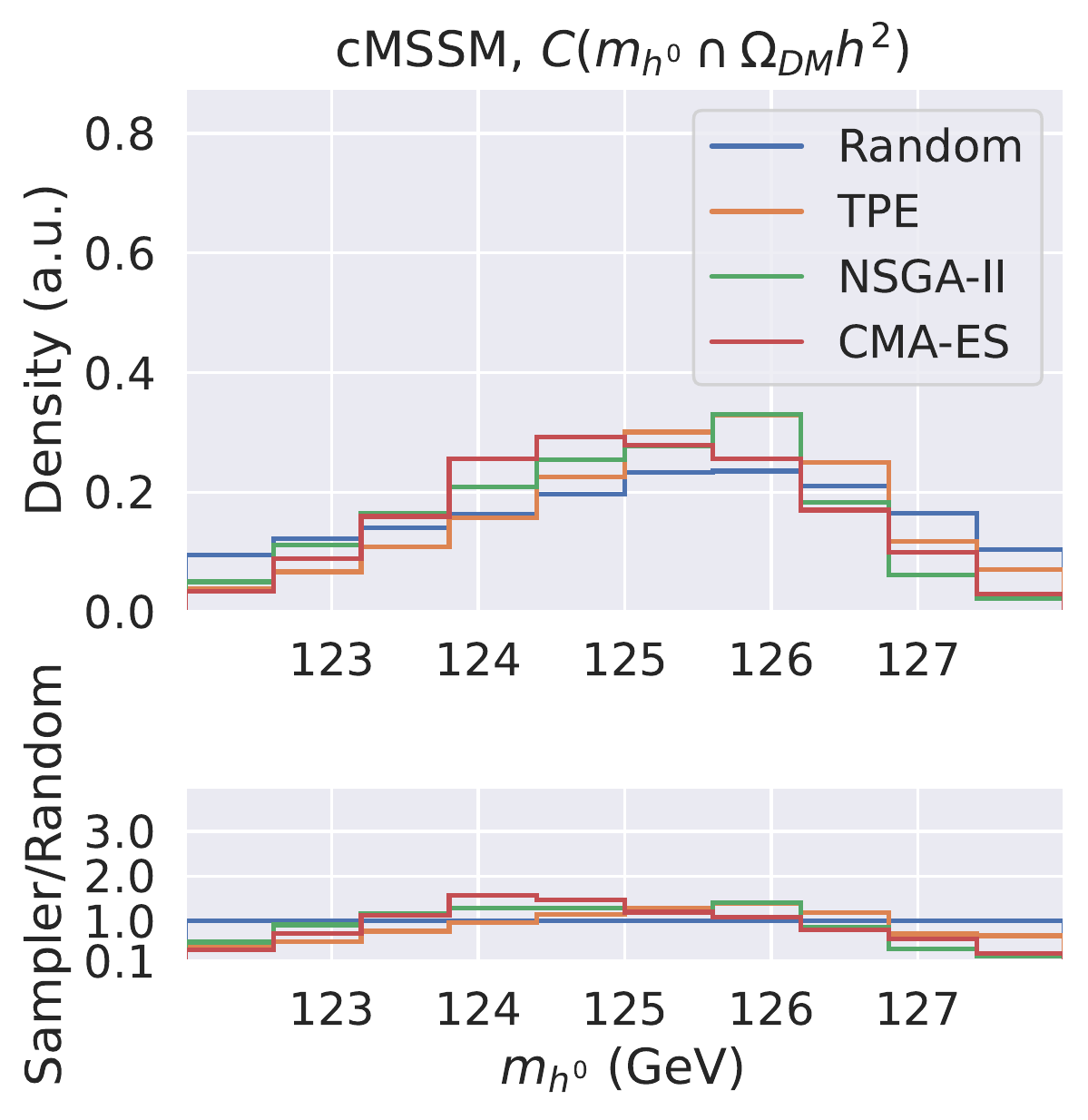}
  \caption{Higgs Mass distribution for the scan with the Higgs mass and dark matter relic density constraints.}
  \label{fig:cMSSM_mh0_with_mo}
\end{subfigure}\\
\begin{subfigure}{.4\textwidth}
  \centering
  \includegraphics[width=1.0\linewidth]{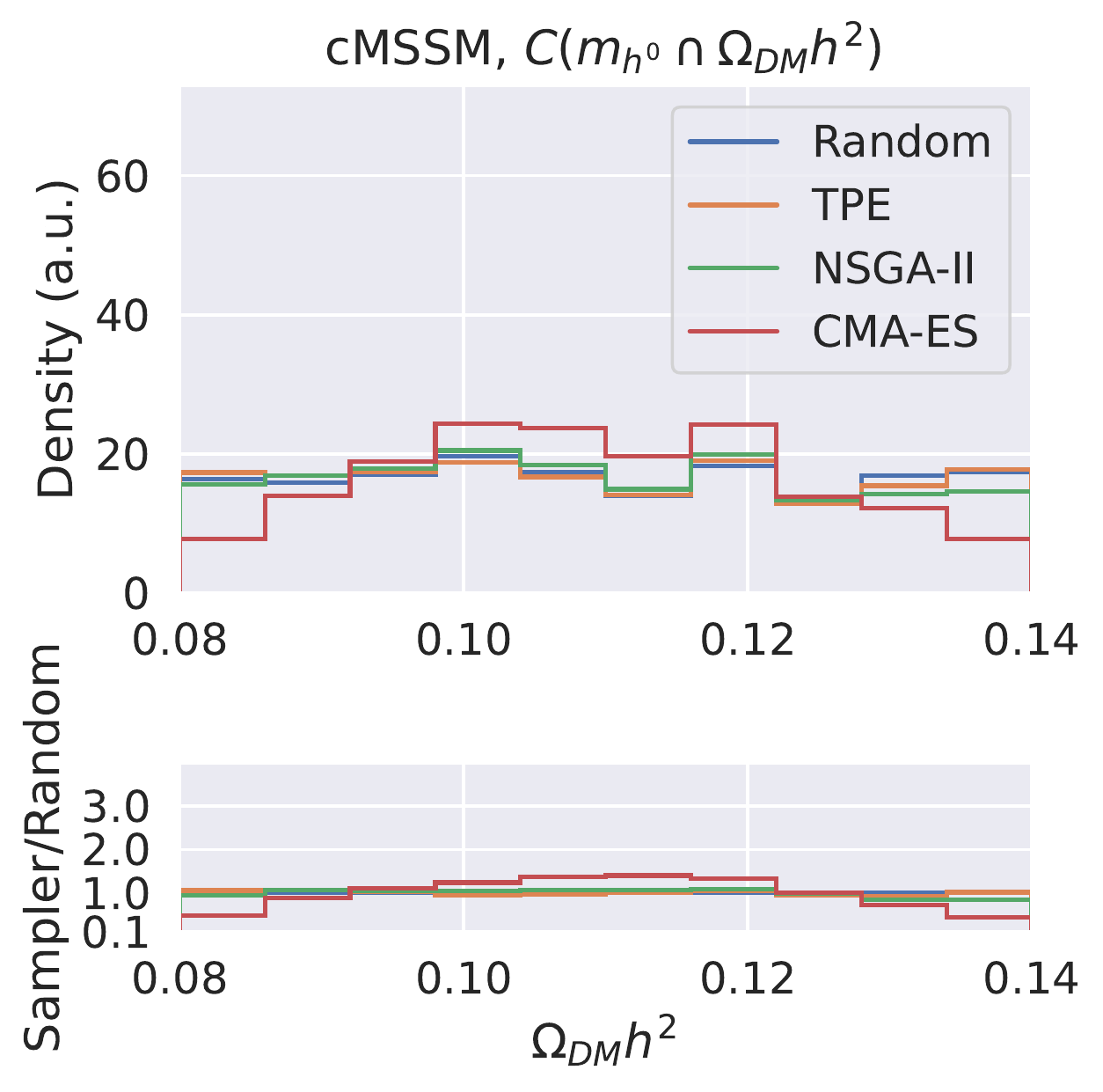}
  \caption{Dark matter relic density distribution for the scan with the Higgs mass and dark matter relic density constraints.}
  \label{fig:cMSSM_omega_with_mo}
\end{subfigure}
\caption{Top panels: Target observables distributions for the cMSSM scans. The resulting valid points histograms for each sampler are produced from joining all the episodes. Bottom panels: The ratio between the histogram of the random sampler with the remaining samplers. In all cases the histograms represent a density, which the area equals to one.}
\label{fig:cMSSM_observables}
\end{figure}

In~\cref{fig:cMSSM_observables} we can see the distributions for the Higgs mass and the dark matter relic density for the cMSSM for each sampler, in the top panels. In the bottom panels we show the ratio of the histogram of each sampler against the random sampler to further illustrate how different samplers produce different distributions.

We notice that TPE and NSGA-II both produce distributions relatively close to the random sampler ones, while CMA-ES exhibits more pronounced deformations. In more detail, we see how CMA-ES seems to centre the distributions far closer to the centre of the allowed interval for each observable.

\begin{figure}
    \centering
    \includegraphics[width=0.7\linewidth]{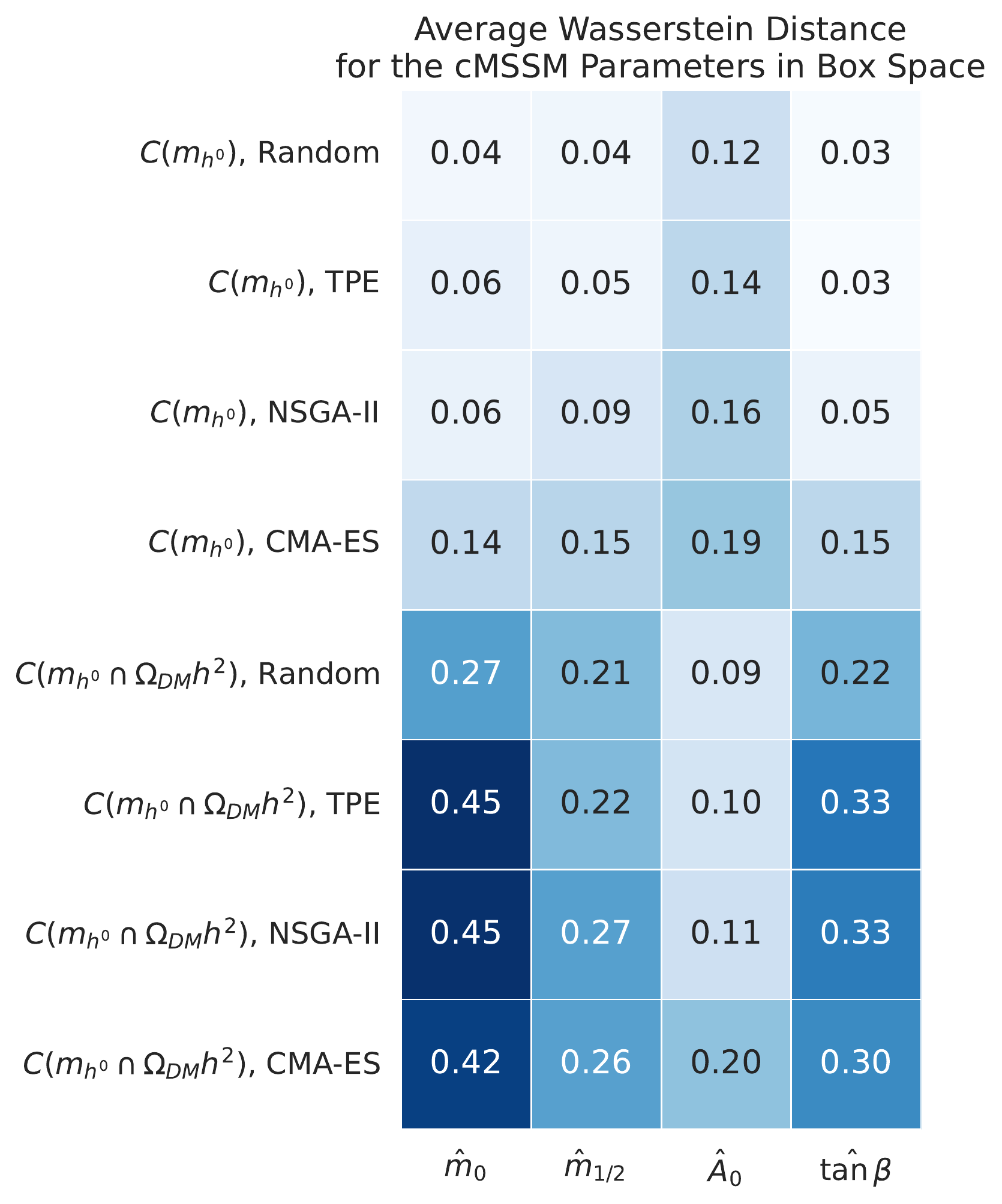}
    \caption{Episode average of the Wasserstein Distance computed on valid points for each (boxed) parameter for each sampler for the cMSSM scans.}
    \label{fig:cMSSM_wd_heatmap}
\end{figure}

In~\cref{fig:cMSSM_wd_heatmap} we present the average over episodes of the Wasserstein distance for each distribution. This measures how much  the distributions of valid points differ to the uniform distribution, as to quantify the parameter space coverage of each sampler. Smaller (larger) values of the Wasserstein distance mean that the distribution is more similar (different) to a uniform distribution. 

As expected, the random sampler is the one that is closest to produce uniform distributions for the parameters, where the deviations between the resulting parameter distributions from the uniform distributions result from the constraint functions. For the other samplers, the higher values of the Wasserstein distance is a result of the sampling algorithm, given the differences in the way each sampler dynamically looks for valid points. We note that for the scan constrained only by the Higgs mass, the CMA-ES sampler considerably distorts the distributions related to $m_0$ and $A_0$ in a far more pronounced manner than the remainder, indicating that it attempts to exploit the relations between these parameters and the Higgs mass. For the case with dark matter relic density constraint, we notice that all non-random samplers noticeably distort the  $m_0$ and $A_0$ distributions,as well as the distribution of $m_{1/2}$, a parameter that directly affects the neutralino mass spectrum and therefore dark matter relic density values.

\begin{figure}
    \centering
    \includegraphics[width=0.8\linewidth]{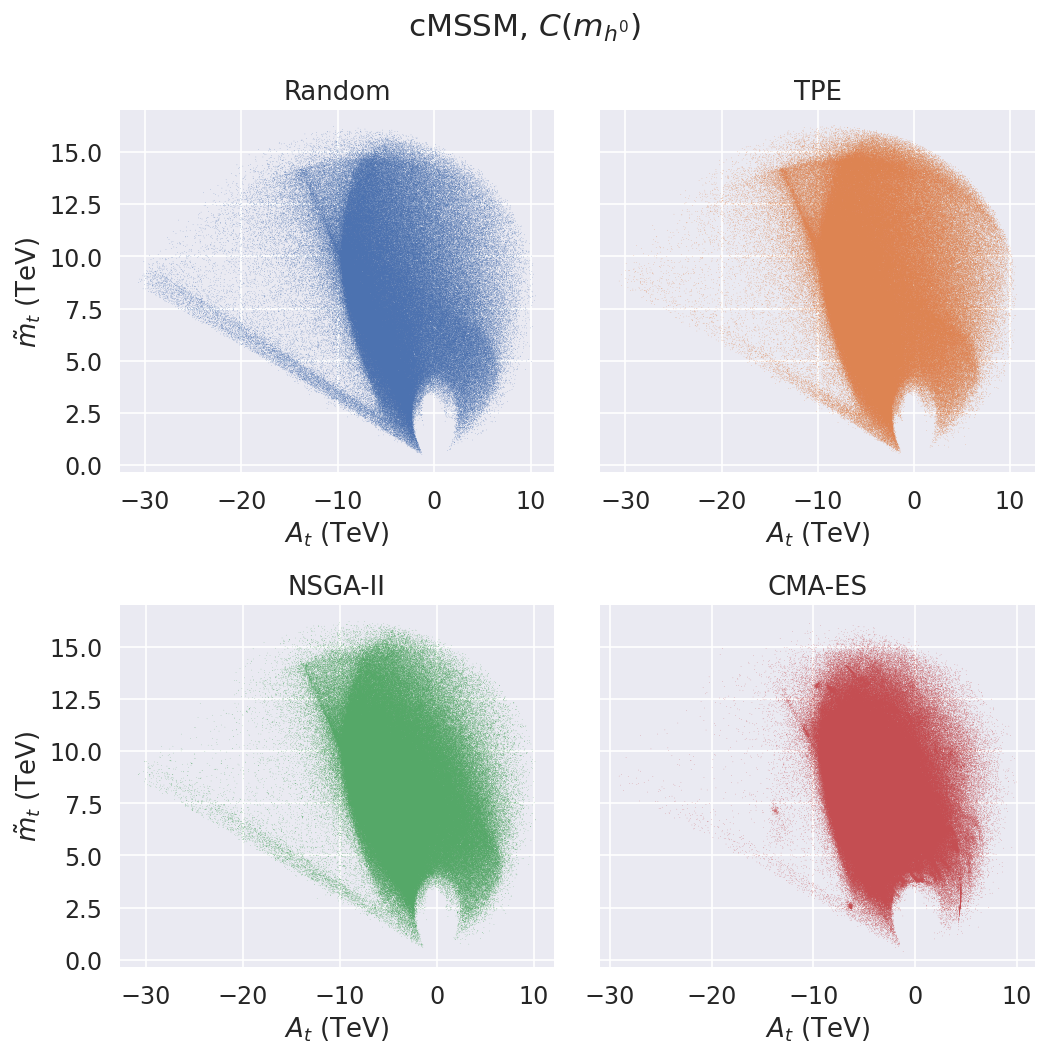}
    \caption{$(\tilde m_t=\sqrt{m_{\tilde t_1} m_{\tilde t_2}}, A_t)$ scatter plots of valid points for the cMSSM scan for each sampler constrained by the Higgs mass.}
    \label{fig:cMSSM_At_vs_mt_wo_mo}
\end{figure}

Another way to look into the differences in parameter distributions across the samplers is to look into scatter plots of relevant pairs of parameters. In~\cref{fig:cMSSM_At_vs_mt_wo_mo} we show the $(\tilde m_t=\sqrt{m_{\tilde t_1} m_{\tilde t_2}}, A_t)$ scatter plot for the cMSSM constrained only by the Higgs mass, which is parametrically dependent on these cMSSM parameters. We observe that the random sampler has the widest area coverage, specially in comparison with CMA-ES, which presents a deficit of points in the $A_t<0$ region. We also notice how the TPE covers the same region with fairly uniform density, whereas NSGA-II was capable of identifying the $\tilde m_t \propto - 1/2 A_t$ region with higher density than the other two non-random samplers. We note for completeness that the reason of the preference of negative values for $A_t$ is pure RGE effect as $A_t \simeq -2 m_{1/2}-0.2 A_0$ for small $\tan\beta$, see e.g.~\cite{Blair:2002pg} and refs.~therein.

\begin{figure}[H]
    \centering
    \includegraphics[width=0.8\linewidth]{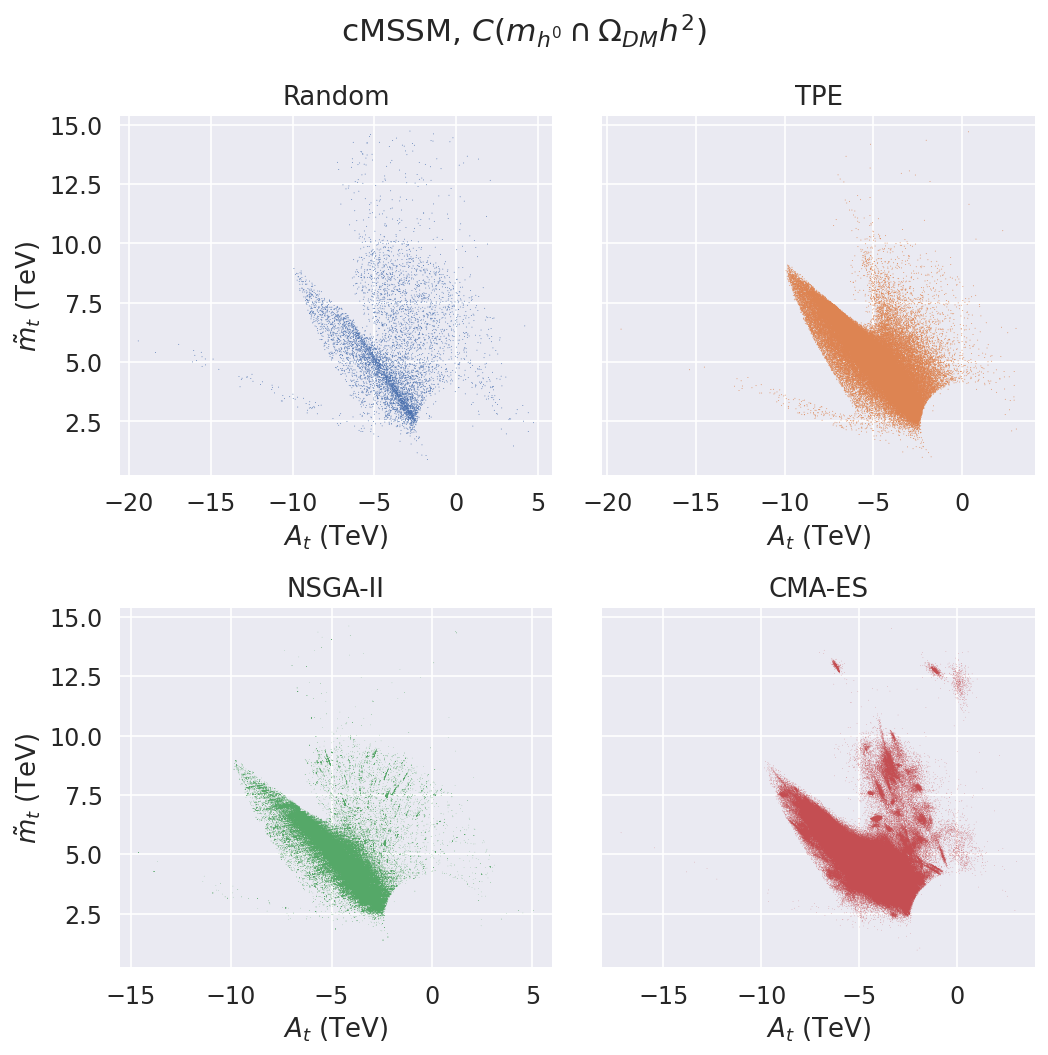}
    \caption{$(\tilde m_t=\sqrt{m_{\tilde t_1} m_{\tilde t_2}}, A_t)$ scatter plots of valid points for the cMSSM scan for each sampler constrained by the Higgs mass and the dark matter relic density.}
    \label{fig:cMSSM_At_vs_mt_with_mo}
\end{figure}

In~\cref{fig:cMSSM_At_vs_mt_with_mo} we can observe how these scatter plots change once we include the dark matter relic density constraint. In this scan, which is far more difficult than the one without this extra constraint, we can observe new features which highlight the differences between the different samplers. Firstly we see that the three non-random samplers produced greater densities in more central regions and seem to fail to produced as many valid points further from these regions. Secondly, we can observe artifacts in the NSGA-II scatter where there is an emerging texture of vertical strips of higher density. This is a known result of genetic algorithms, where new suggested points inherit values from their parents, which can lead to the same value to be reused over many generations\footnote{In the genetic algorithms literature, recurrent combinations that survive through generations are called schema.} producing these strips. Finally, in the CMA-ES we observe many smaller regions of high density, which are explained by the nature of the sampler itself, since it \emph{eagerly} samples from a multivariate normal with rolling statistics of the best points, i.e. it \emph{exploits} the learned statistics of a local population, producing many valid points in the vicinity of the rolling mean of the best points. Due to the \emph{eager} nature of the CMA-ES, we can also observe how it fails to capture smaller regions of valid points away from the \emph{easier} region, while producing highly condensed regions of points where other samplers have only found a few, for example on the upper right quadrant.

\begin{figure}
    \centering
    \includegraphics[width=0.8\linewidth]{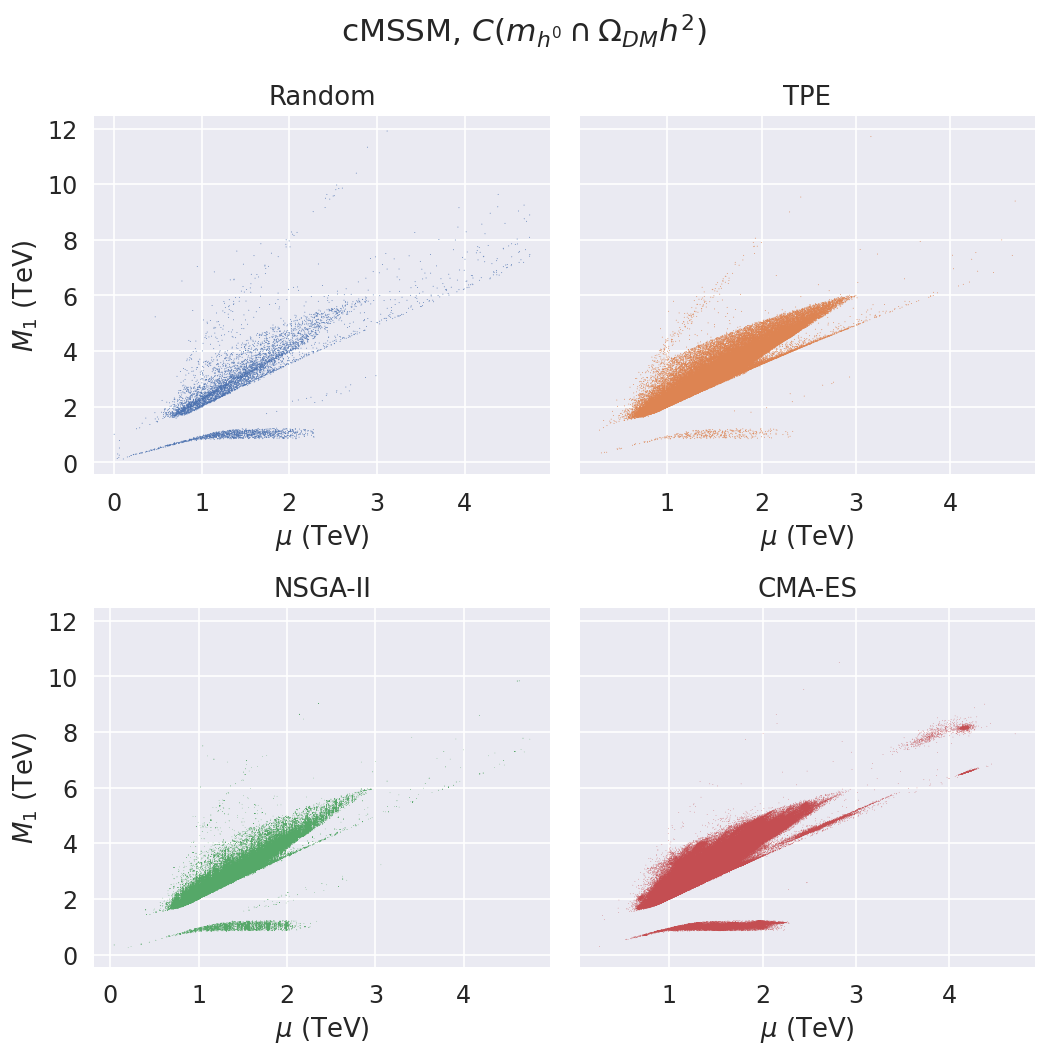}
    \caption{$(\mu, M_1)$ scatter plots of valid points for the cMSSM scan for each sampler constrained by the Higgs mass and the dark matter relic density.}
    \label{fig:cMSSM_m1_vs_mu}
\end{figure}

With the dark matter relic constraint it is informative to look at the $(\mu, M_1)$\footnote{We omit the equivalent scatter with $M_2$ as in the cMSSM $M_1 \sim M_2$ and therefore this plot provides no new insight.} scatter plots as these are the relevant parameters for dark matter phenomenology. These are presented in~\cref{fig:cMSSM_m1_vs_mu}. Again, we see how the non-random samplers produce far denser regions of valid points, while still struggling to cover the parameter space the same way as the random sampler. However, both the TPE and the NSGA-II reproduce the overall features of the region obtained by the random sampler, whereas CMA-ES exhibits again its \emph{eager}  nature, \emph{e.g.} we can see small patches of high density arising, while failing to populate the $M_1\gg \mu$ faint region. Interestingly enough, all samplers discovered the $M_1 \lesssim 1$ TeV disconnected region, providing some evidence that these samplers can find multimodal solutions.

\subsubsection{pMSSM}

We now turn to the pMSSM. Given that our pMSSM scan covers 19 parameters, as opposed to the four parameters of the cMSSM, this scan will allow us to study the impact of increasing the dimensionality of the parameter space in the performance and results of different samplers.

\begin{figure}
\centering
\begin{subfigure}{.4\textwidth}
  \centering
  \includegraphics[width=1.0\linewidth]{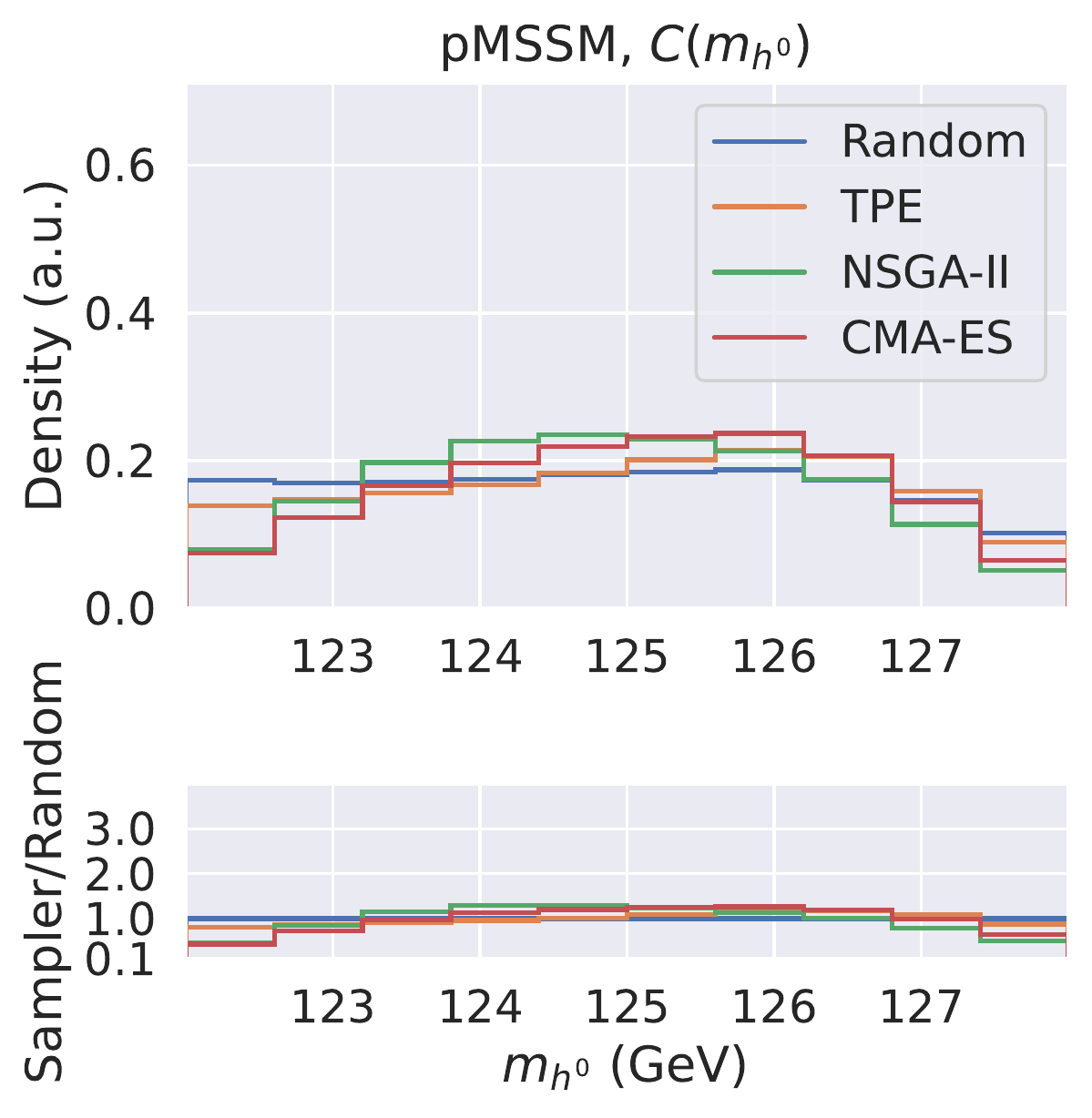}
   \caption{Higgs Mass distribution for the scan with
the Higgs mass constraint.}
  \label{fig:pMSSM_mh0_wo_mo}
\end{subfigure}%
\begin{subfigure}{.4\textwidth}
  \centering
  \includegraphics[width=1.0\linewidth]{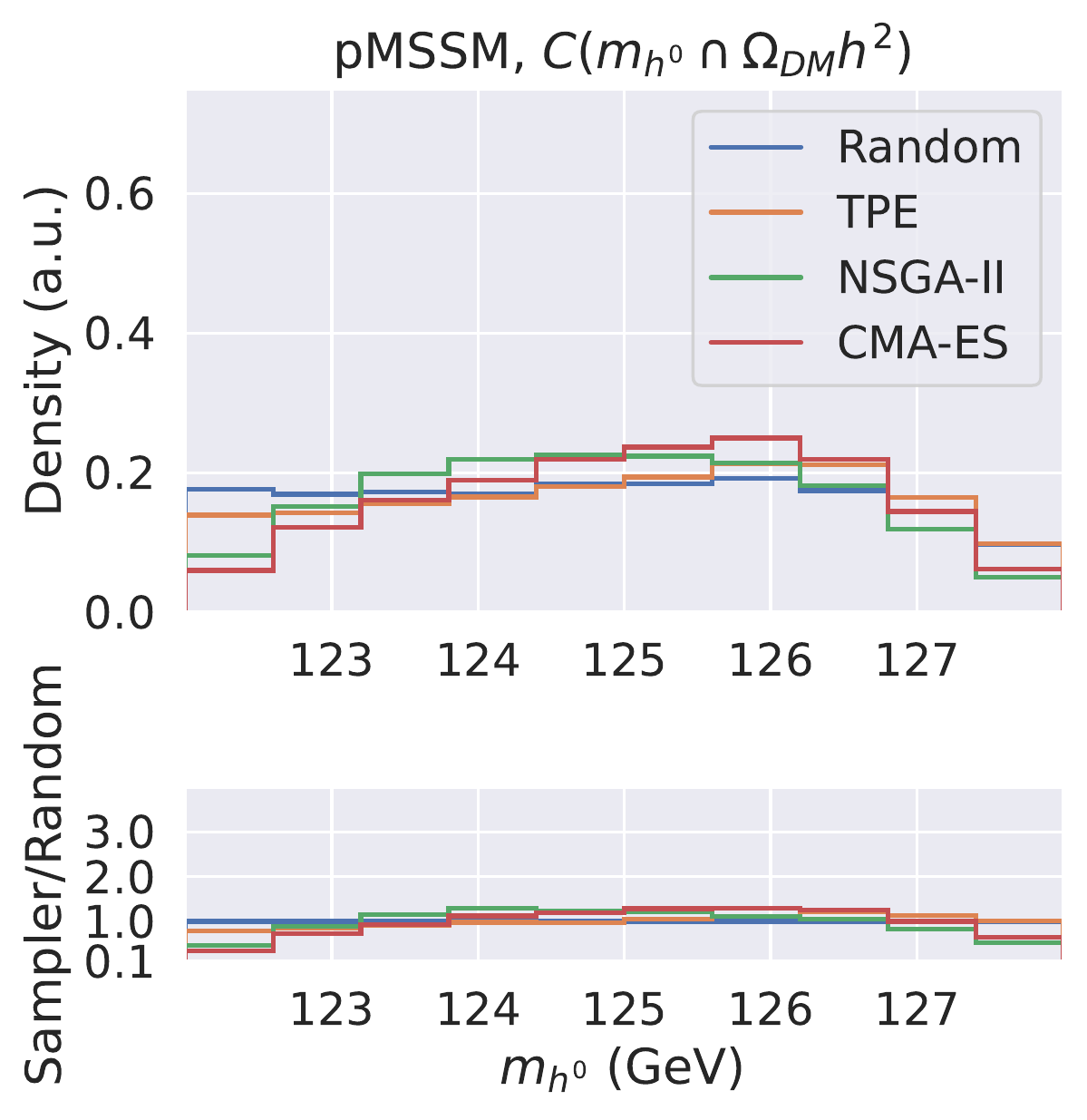}
  \caption{Higgs Mass distribution for the scan with the Higgs mass and dark matter relic density
constraints.}
  \label{fig:pMSSM_mh0_with_mo}
\end{subfigure}\\
\begin{subfigure}{.4\textwidth}
  \centering
  \includegraphics[width=1.0\linewidth]{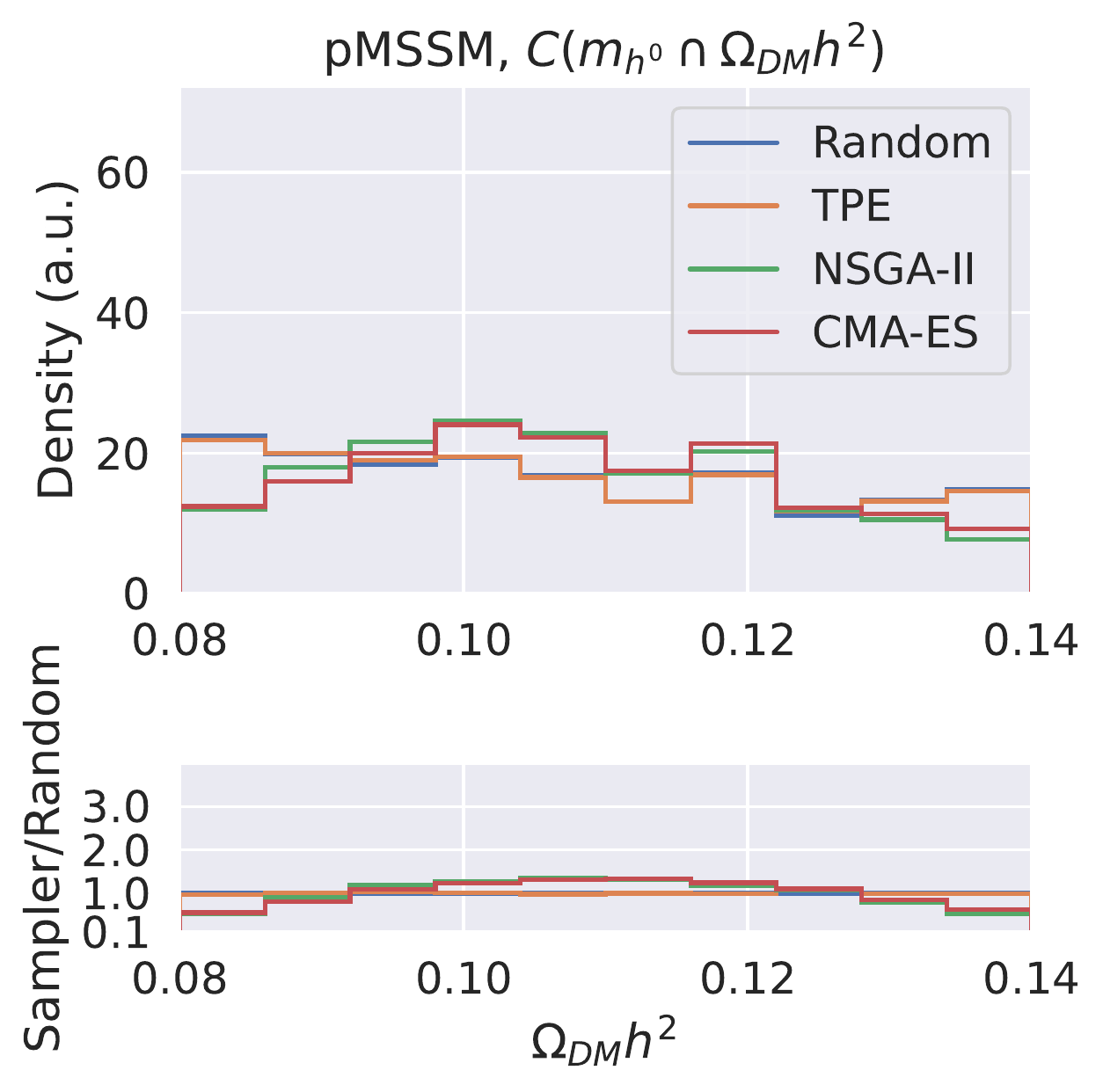}
  \caption{Dark matter relic density distribution for the scan with the Higgs mass and dark matter relic density constraints.}
  \label{fig:pMSSM_omega_with_mo}
\end{subfigure}
\caption{Top panels: Target observables distributions for the pMSSM scans. The resulting valid points
histograms for each sampler are produced from joining all the episodes. Bottom panels: The
ratio between the histogram of the random sampler with the remaining samplers.  In all cases the histograms represent a density, which the area equals to one.}
\label{fig:pMSSM_observables}
\end{figure}

In~\cref{fig:pMSSM_observables} we present the resulting distributions for the Higgs mass and the dark matter relic density for both pMSSM scans. Similarly to the cMSSM scans, the non-random samplers focus their valid points in the interior region of the allowed interval for each observable, with the TPE being the sampler that produces distributions more similar to the random sampler.

\begin{figure}
    \centering
    \includegraphics[width=1.0\linewidth]{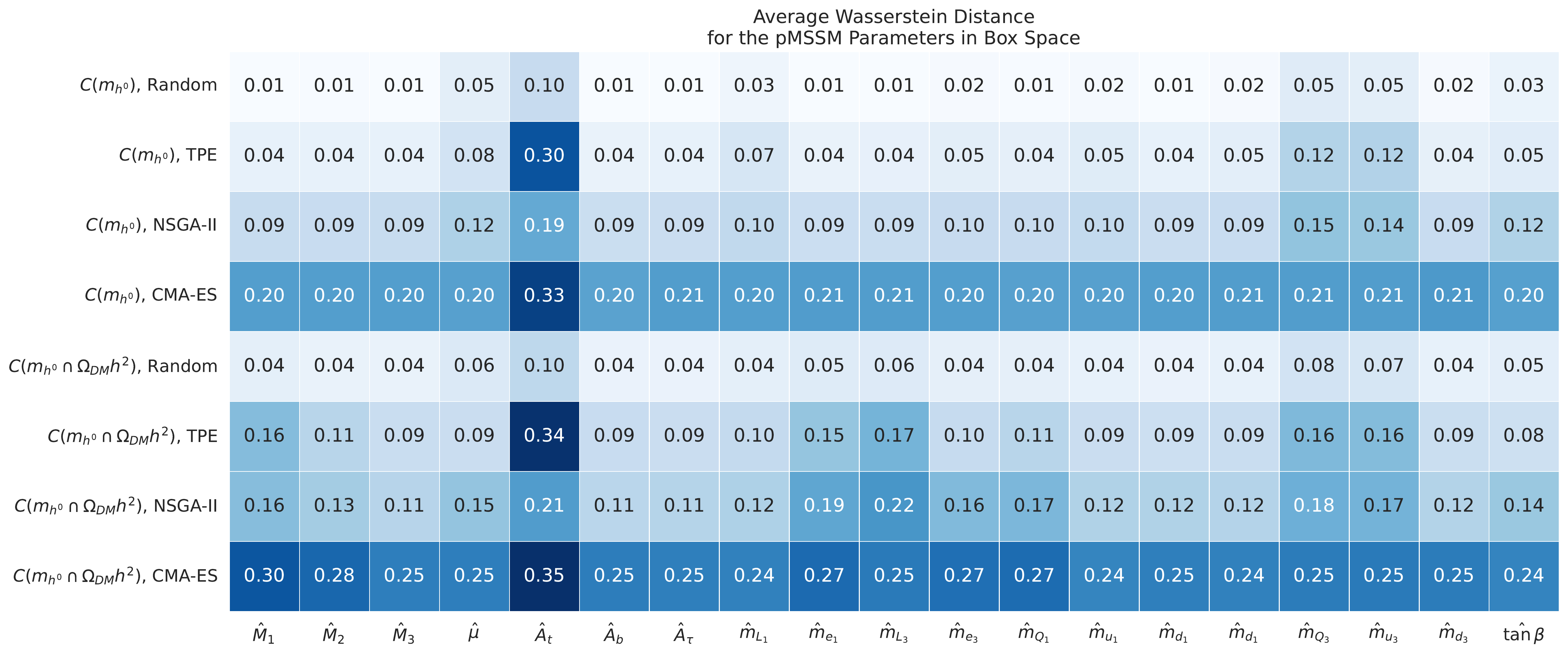}
    \caption{Episode average of the Wasserstein Distance computed on valid points for each (boxed) parameter for each sampler for the pMSSM scans.}
    \label{fig:pMSSM_wd_heatmap}
\end{figure}

As with the cMSSM, we omit the distributions of the parameters in this section for the sake of a light discussion and instead we present the episode average Wasserstein distance for the parameter sof the pMSSM scans in~\cref{fig:pMSSM_wd_heatmap}. The distributions for all pMSSM parameters can found in the git code repository.

Just like in the cMSSM case, the random sampler produces the smallest deviations from the uniform distributions, due to its unmodified sampling. Next, we see that TPE produces almost no further distortions in the parameters, except for those directly related to the Higgs mass -- $A_t$, $\tilde m_{Q_3}$, $\tilde m_{u_3}$ -- for the scan without the dark matter relic density constraint. When switching on the dark matter relic density constraint, the TPE produces further distortions in the parameters associated with dark matter phenomenology, namely $M_1$, $M_2$, $\mu$.
In addition also the slepton mass parameters
are distorted as the co-annihilation channels
become important if the mass difference between sleptons and neutralinos becomes sufficiently small and if the lightest neutralino has a sizeable bino-component. Similarly the enhanced distortion for the third generation squarks occurs due to the part of parameter region where there is a stop-neutralino co-annihilation if the lightest neutralino has a sizeable higgsino component. 
Unsurprisingly, the CMA-ES is the sampler that produces the most different parameter distributions due to its \emph{eager} nature. The fact that the TPE does not distort the distributions more is somehow surprising, as it makes use of Gaussian Mixture Models, a density learning algorithm that can be prone to the curse of dimensionality, whereas genetic algorithms such as NSGA-II are robust against this problem as they are not reliant on a learnable model. 

\begin{figure}
    \centering
    \includegraphics[width=0.75\linewidth]{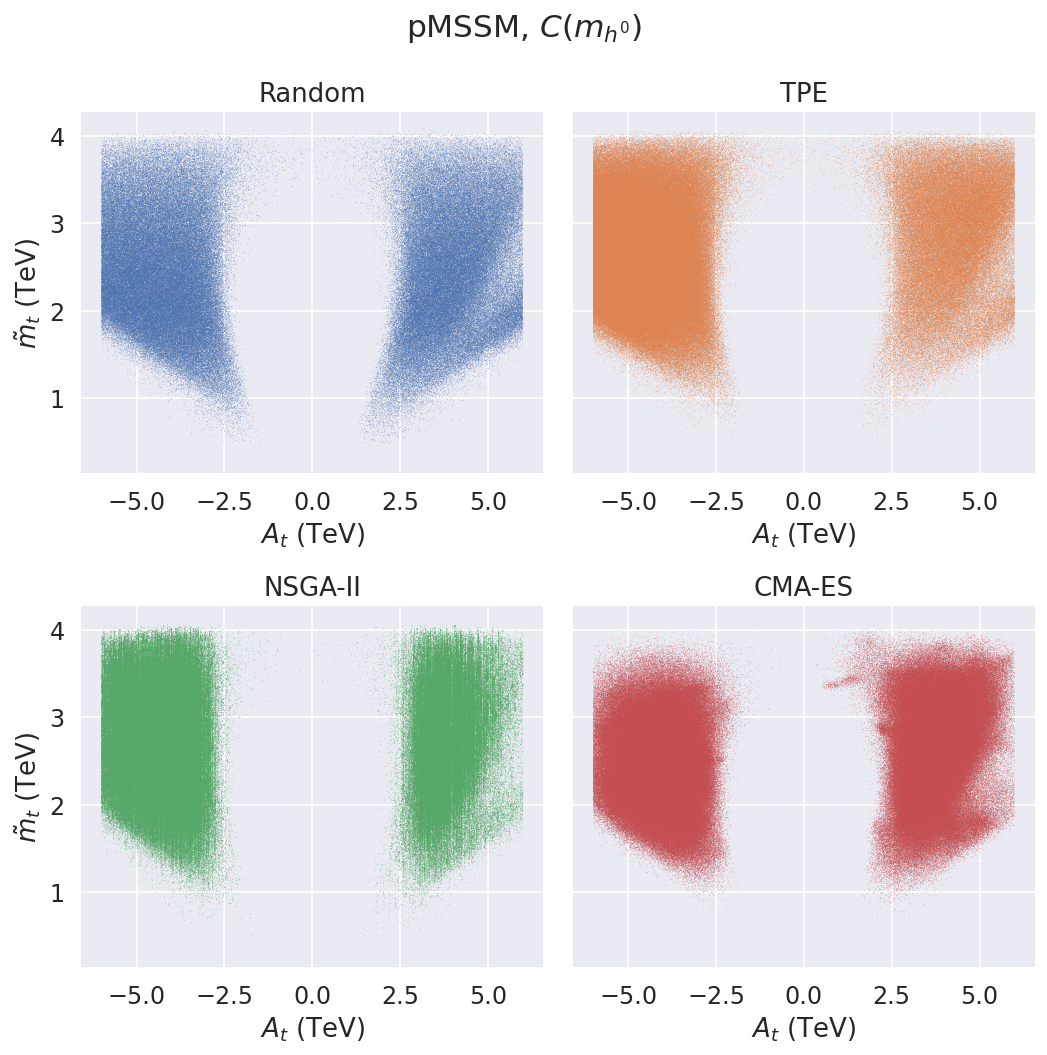}
    \caption{$(A_t, \tilde m_t)$ scatter plot of valid points for the pMSSM scan for each sampling algorithm constrained by the Higgs mass.}
    \label{fig:pMSSM_At_vs_mt_with_mo}
\end{figure}

We further investigate the impact that different samplers can have on the parameter distributions by looking at a selection of scatter plots. In~\cref{fig:pMSSM_At_vs_mt_with_mo} we present the $(A_t, \tilde m_t)$ scatter plot for the pMSSM scan constrained by the Higgs mass, where we can see that TPE is covering the same region as the random sampler with fairly constant point density. Furthermore, we can identify once again NSGA-II artifacts by noticing the emergence of strips of higher density, associated with the nature of how genetic algorithms produce new points via offspring. Finally, we see how CMA-ES focuses on easy regions of the parameter space due to its \emph{eager} nature, more concretely we notice how it produces far less points in regions of small $\tilde m_t$ in comparison to the other samplers.

\begin{figure}
    \centering
    \includegraphics[width=0.75\linewidth]{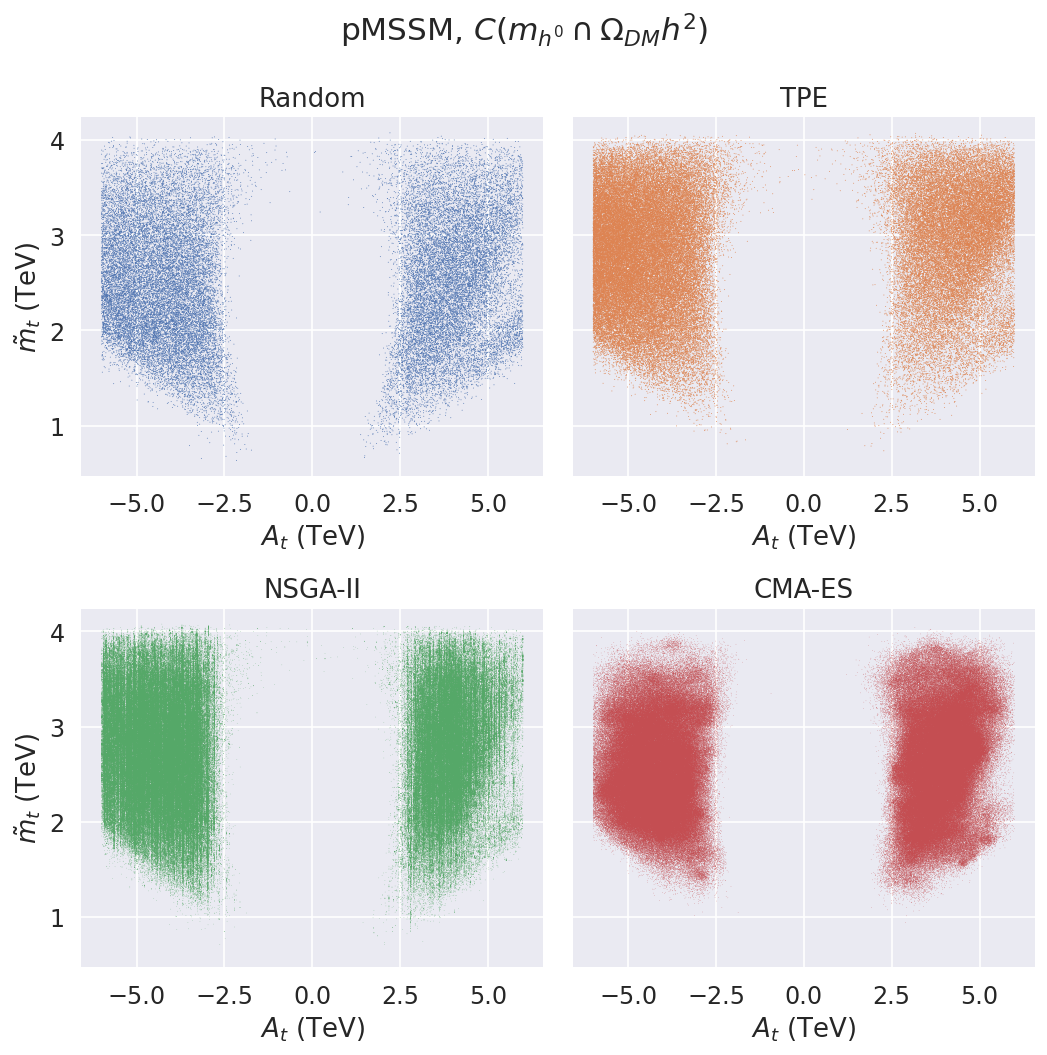}
    \caption{$(A_t, \tilde m_t)$ scatter plot of valid points for the pMSSM scan for each sampling algorithm constrained by the Higgs mass and the dark matter relic density.}
    \label{fig:pMSSM_At_vs_mt_wo_mo}
\end{figure}

Looking at the equivalent scatter plots for the scan with the dark matter relic density included in~\cref{fig:pMSSM_At_vs_mt_wo_mo}, we observe similar features and behaviours. With special highlights to how the CMA-ES presents again smaller oval regions of higher density and the clear strips of higher density in the NSGA-II scatter, while the TPE produces a very similar result to the random sampler.

\begin{figure}
    \centering
    \includegraphics[width=0.75\linewidth]{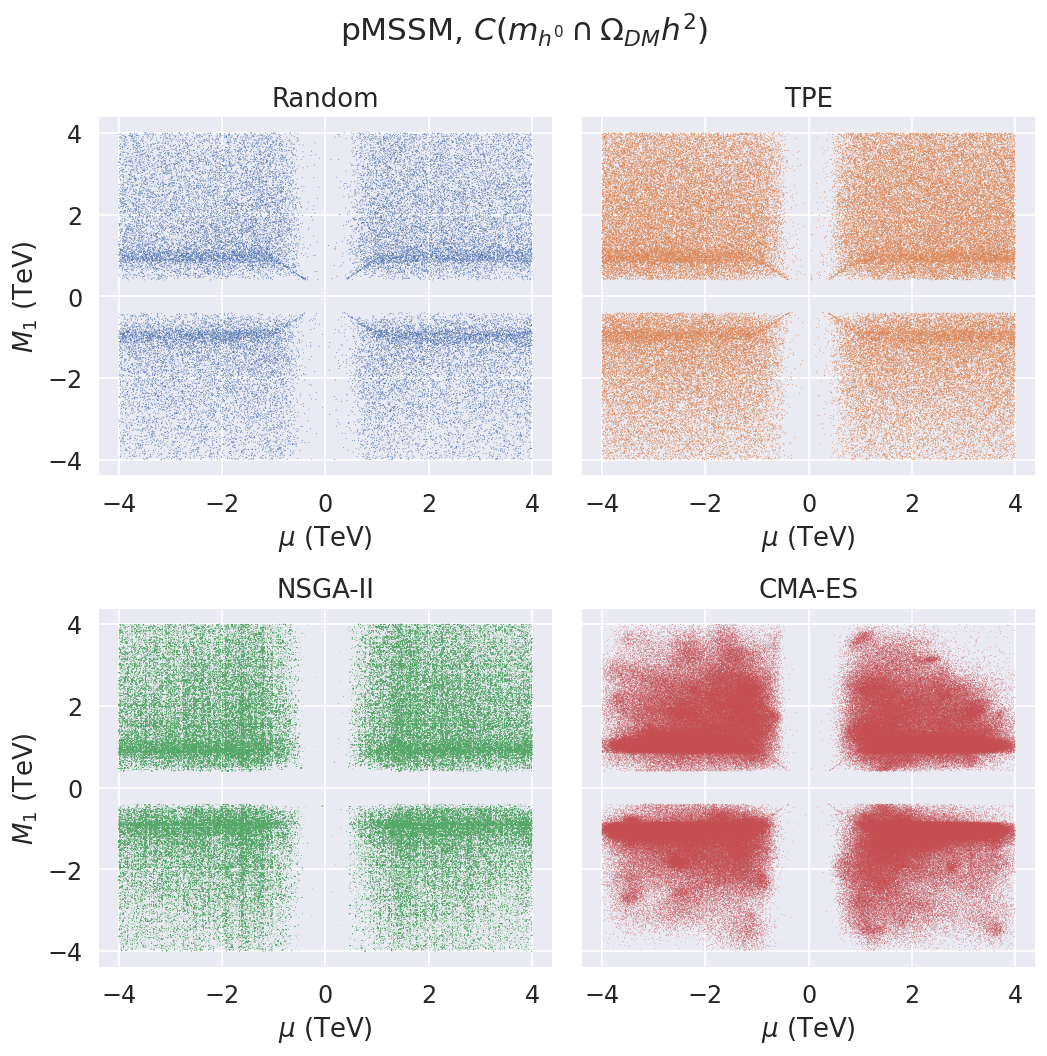}
    \caption{$(\mu, M_1)$ scatter plot of valid points for the pMSSM scan for each sampler constrained by the Higgs mass and the dark matter relic density.}
    \label{fig:pMSSM_m1_vs_mu}
\end{figure}

\begin{figure}
    \centering
    \includegraphics[width=0.75\linewidth]{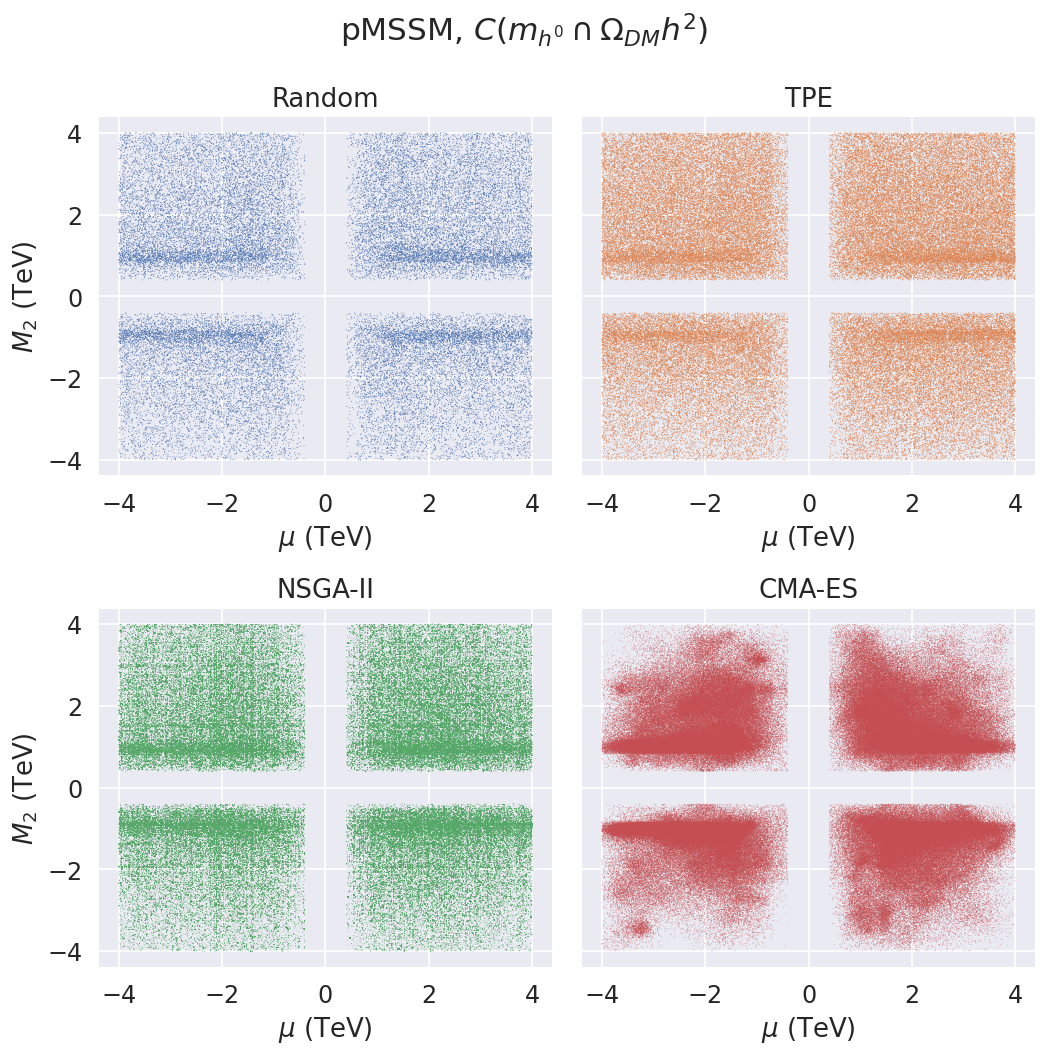}
    \caption{$(\mu, M_2)$ scatter plot of valid points for the pMSSM scan for each sampler constrained by the Higgs mass and the dark matter relic density.}
    \label{fig:pMSSM_m2_vs_mu}
\end{figure}

Continuing the discussion of the pMSSM with dark matter relic density constraints, we now focus on the $(\mu, M_1)$ and $(\mu, M_2)$ scatter plots in~\cref{fig:pMSSM_m1_vs_mu,fig:pMSSM_m2_vs_mu}. Some interesting features emerge in these scatter plots. We notice how the TPE is very similar to the random sampler, including the slightly higher density regions of $| M_1| \gtrsim 1$ TeV and $|M_2|\gtrsim 1$ TeV. We also observe the high density strips artifacts in the NSGA-II scatters, whereas CMA-ES does not cover the same space as the other samplers, and produces patchy regions of higher density. Interestingly, the TPE is able to discover the $|\mu| \sim |M_1|$ regions, which the other non-random samplers seemingly struggle to find.

\subsection{Efficiency and Sampling Metrics}

Having discussed the impact of each sampler in the final parameter distributions in the previous section, in this section we compare the different samplers with respect to their efficiency and other sampling metrics. 

\begin{figure}
\centering
\begin{subfigure}{.45\textwidth}
  \centering
  \includegraphics[width=1.0\linewidth]{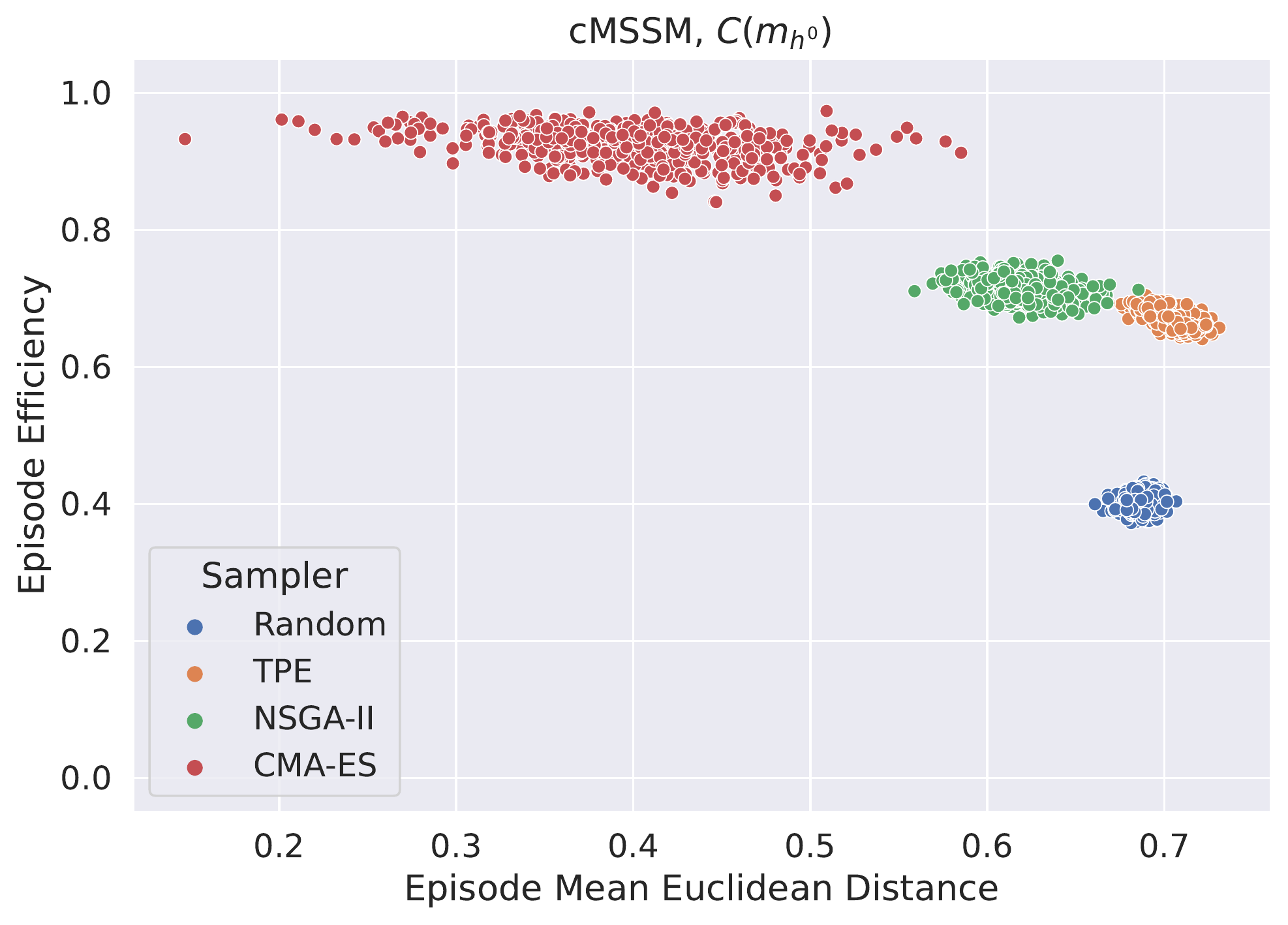}
   \caption{Mean Euclidean distance vs efficiency per episode, for the cMSSM scan with Higgs mass constraint.}
  \label{fig:cMSSM_wo_mo_euclidean_efficiency}
\end{subfigure}%
\begin{subfigure}{.45\textwidth}
  \centering
  \includegraphics[width=1.0\linewidth]{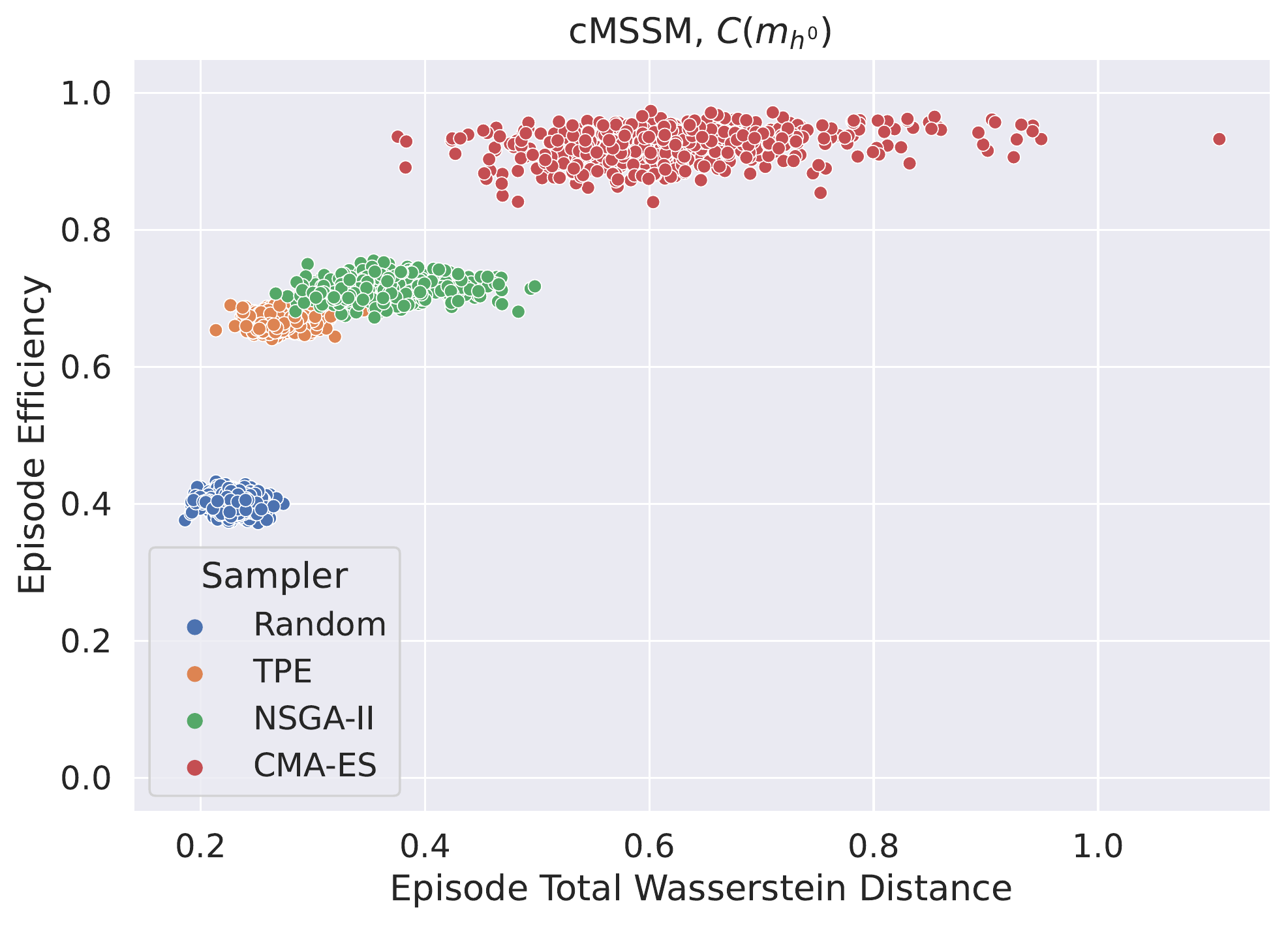}
  \caption{Wasserstein distance vs efficiency per episode, for the cMSSM scan with Higgs mass constraint.}
  \label{fig:cMSSM_wo_mo_wd_efficiency}
\end{subfigure} \\
\begin{subfigure}{.45\textwidth}
  \centering
  \includegraphics[width=1.0\linewidth]{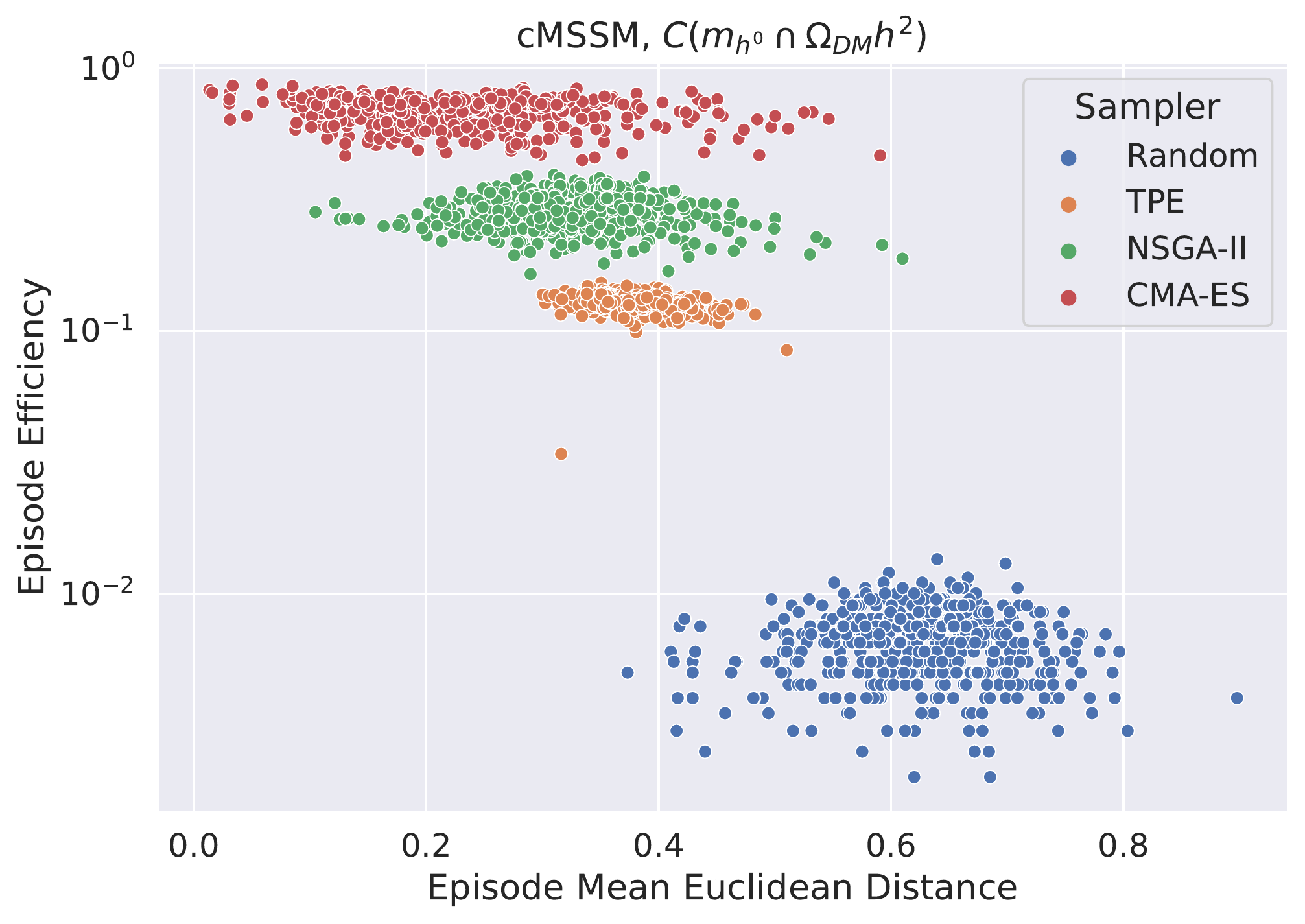}
   \caption{Mean Euclidean distance vs efficiency per episode, for the cMSSM scan with Higgs mass and dark matter relic density constraints.}
  \label{fig:cMSSM_with_mo_euclidean_efficiency}
\end{subfigure}%
\begin{subfigure}{.45\textwidth}
  \centering
  \includegraphics[width=1.0\linewidth]{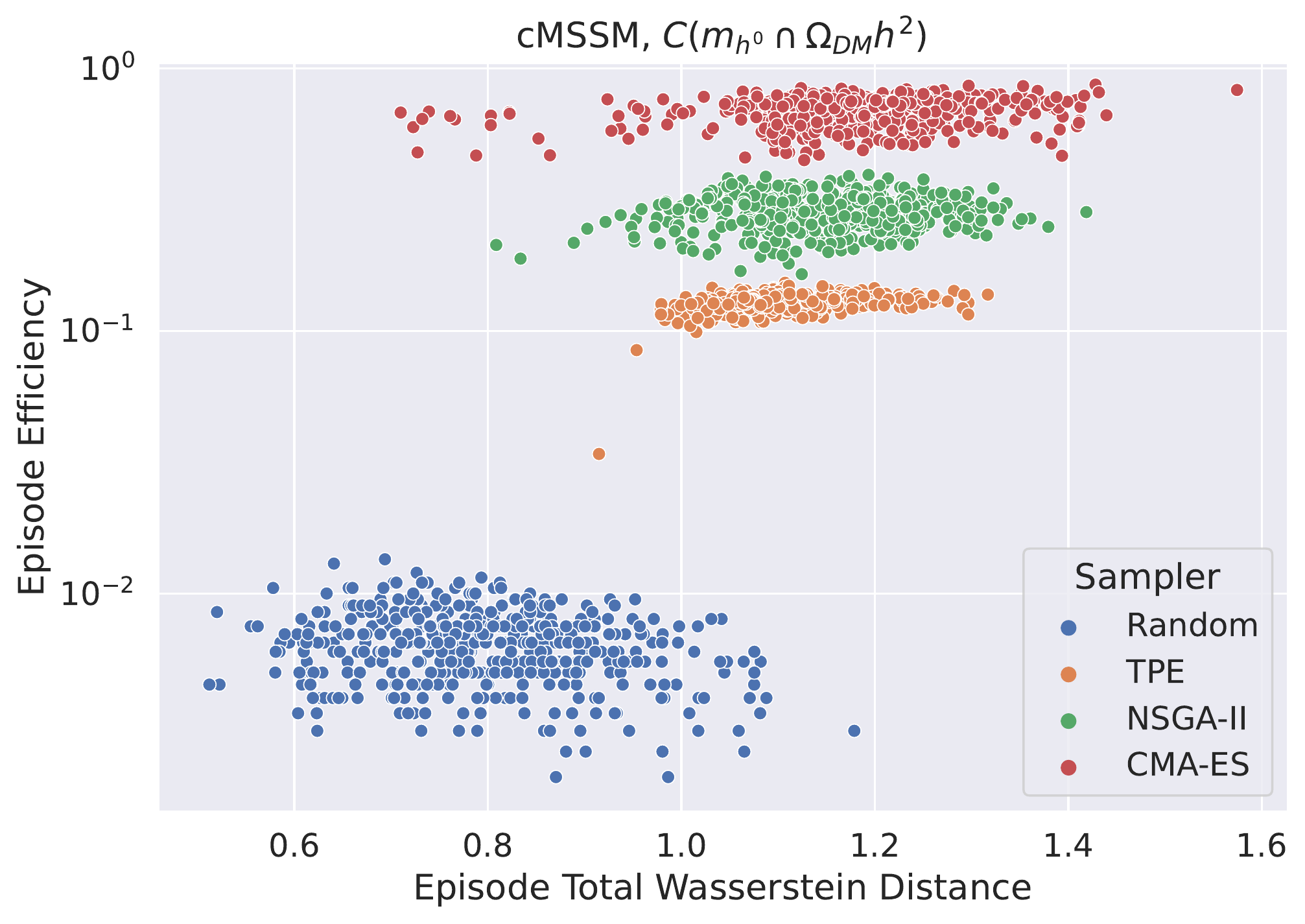}
  \caption{Wasserstein distance vs efficiency per episode, for the cMSSM scan with Higgs mass and dark matter relic density constraints.}
  \label{fig:cMSSM_with_mo_wd_efficiency}
\end{subfigure}
\caption{Efficiency vs Distance metrics, computed using valid points, scatter plots for each sampler for the cMSSM scans.}
\label{fig:cMSSM_scatterplots}
\end{figure}

In~\cref{fig:cMSSM_scatterplots} we can see the scatter plots for the cMSSM scans, with and without the dark matter relic density constraint, for both efficiency vs episode mean euclidean distance and efficiency vs episode total -- i.e., summed over all the parameters -- Wasserstein distance. These highlight the \emph{exploration-exploitation} trade-off, as the most efficient sampler, CMA-ES, provides the worst distance metrics in accordance to the discussion from the previous section. In general, TPE provides the best parameter space coverage with slight less efficiency than the NSGA-II, which produces points which are more clustered together, as we have seen before with the strips of higher density. The NSGA-II episodes have a wider spread of possible values for the Wasserstein distance, providing a good trade-off between coverage and efficiency. We also notice that for the dark matter relic density scan, we gain at least a factor of 10 in parameter sampling efficiency, with CMA-ES increasing efficiency even further. Interestingly, we observe that for that for the cMSSM without Dark Matter relic density constraint, TPE produces on average higher episode mean euclidean distances. This might indicate that TPE, which makes use of clustering points via a Guassian Mixture Model, is sampling from far disjoint patches of the parameter space, increasing the mean euclidean distance within the episodes. This indicates that episode mean euclidean distance might not always be the appropriate metric for parameter space coverage.

\begin{figure}
\centering
\begin{subfigure}{.45\textwidth}
  \centering
  \includegraphics[width=1.0\linewidth]{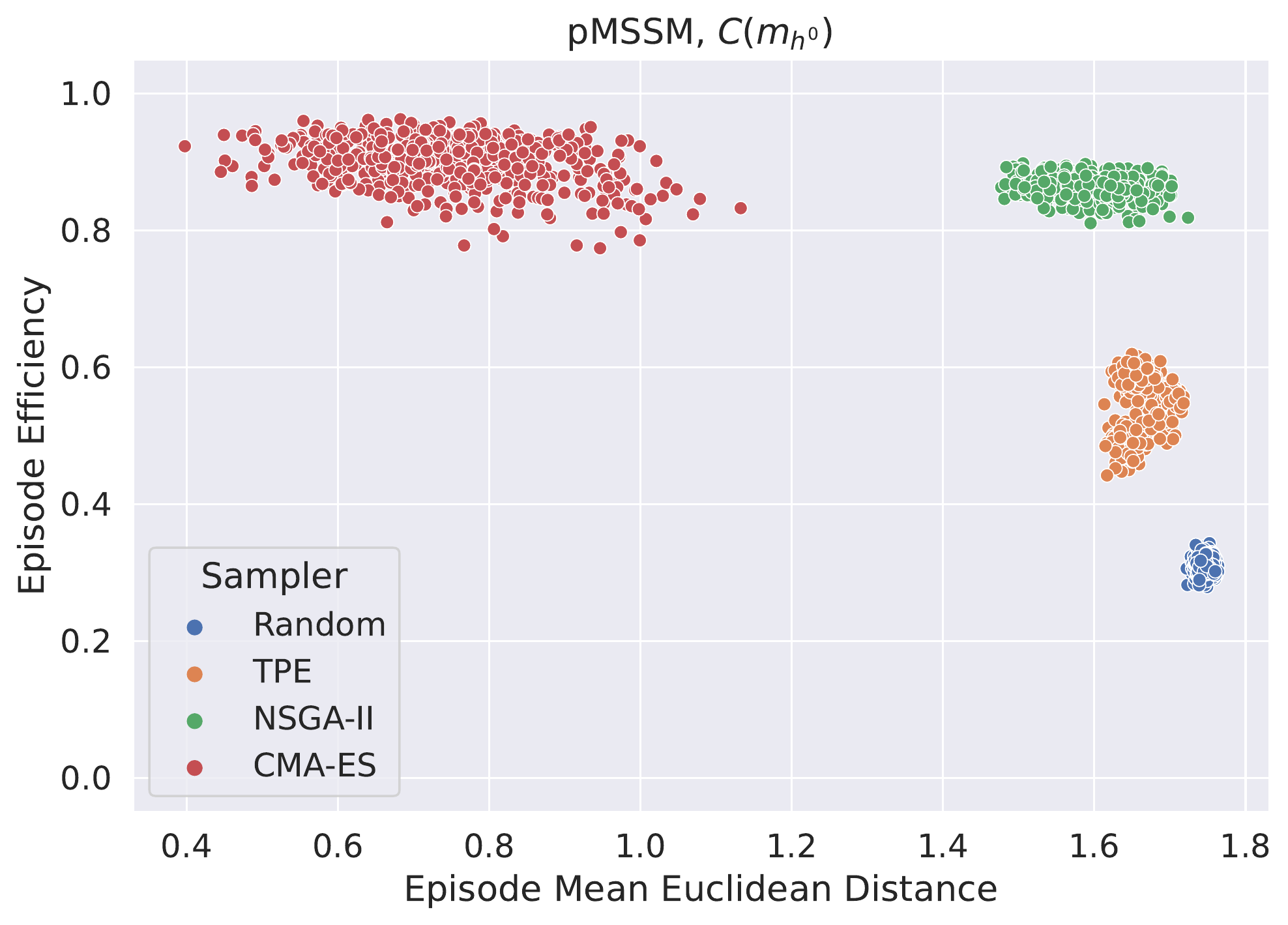}
   \caption{Mean Euclidean distance vs efficiency per episode, for the pMSSM scan with Higgs mass constraint.}
  \label{fig:pMSSM_wo_mo_euclidean_efficiency}
\end{subfigure}%
\begin{subfigure}{.45\textwidth}
  \centering
  \includegraphics[width=1.0\linewidth]{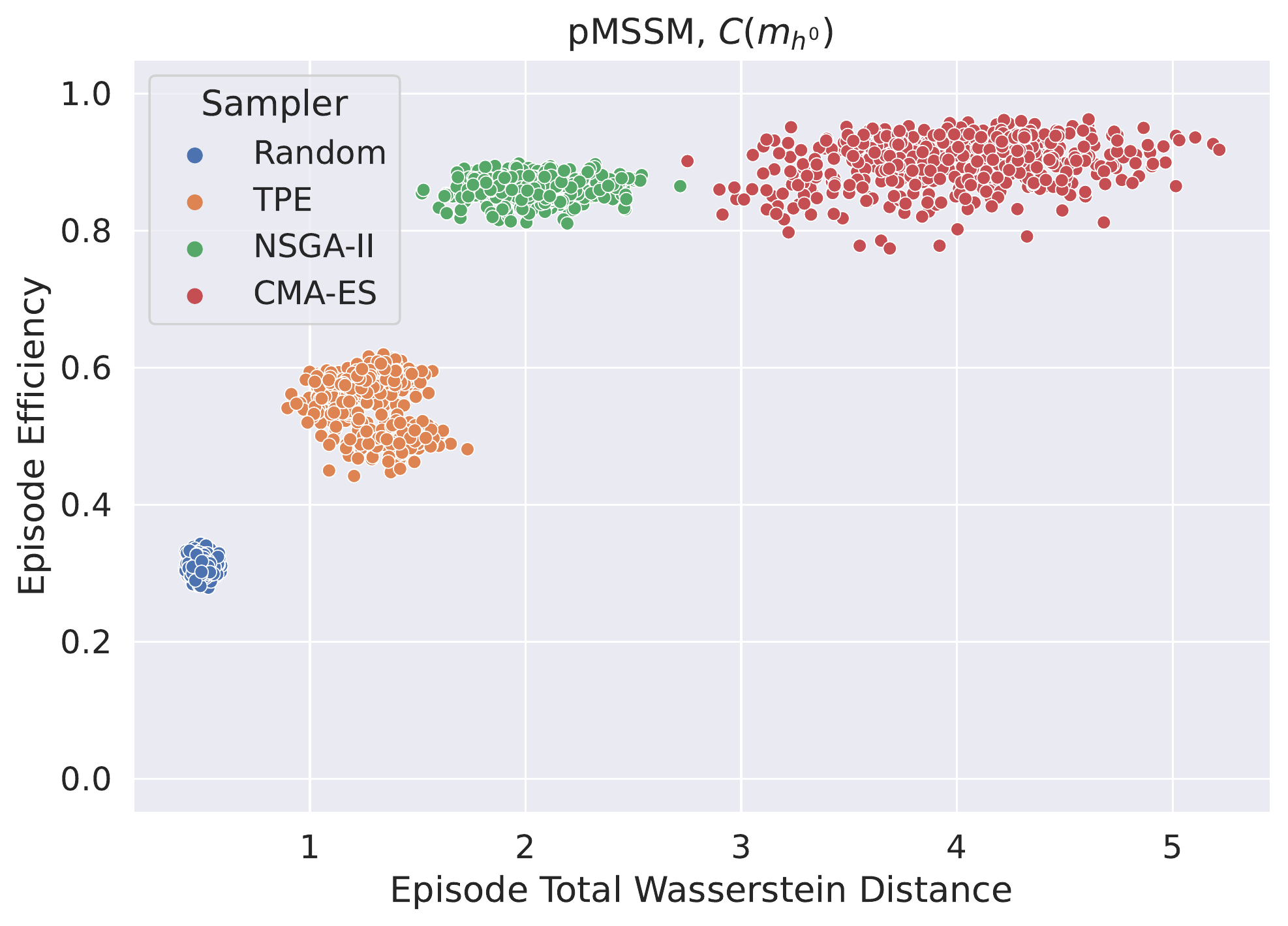}
  \caption{Wasserstein distance vs efficiency per episode, for the pMSSM scan with Higgs mass constraint.}
  \label{fig:pMSSM_wo_mo_wd_efficiency}
\end{subfigure} \\
\begin{subfigure}{.45\textwidth}
  \centering
  \includegraphics[width=1.0\linewidth]{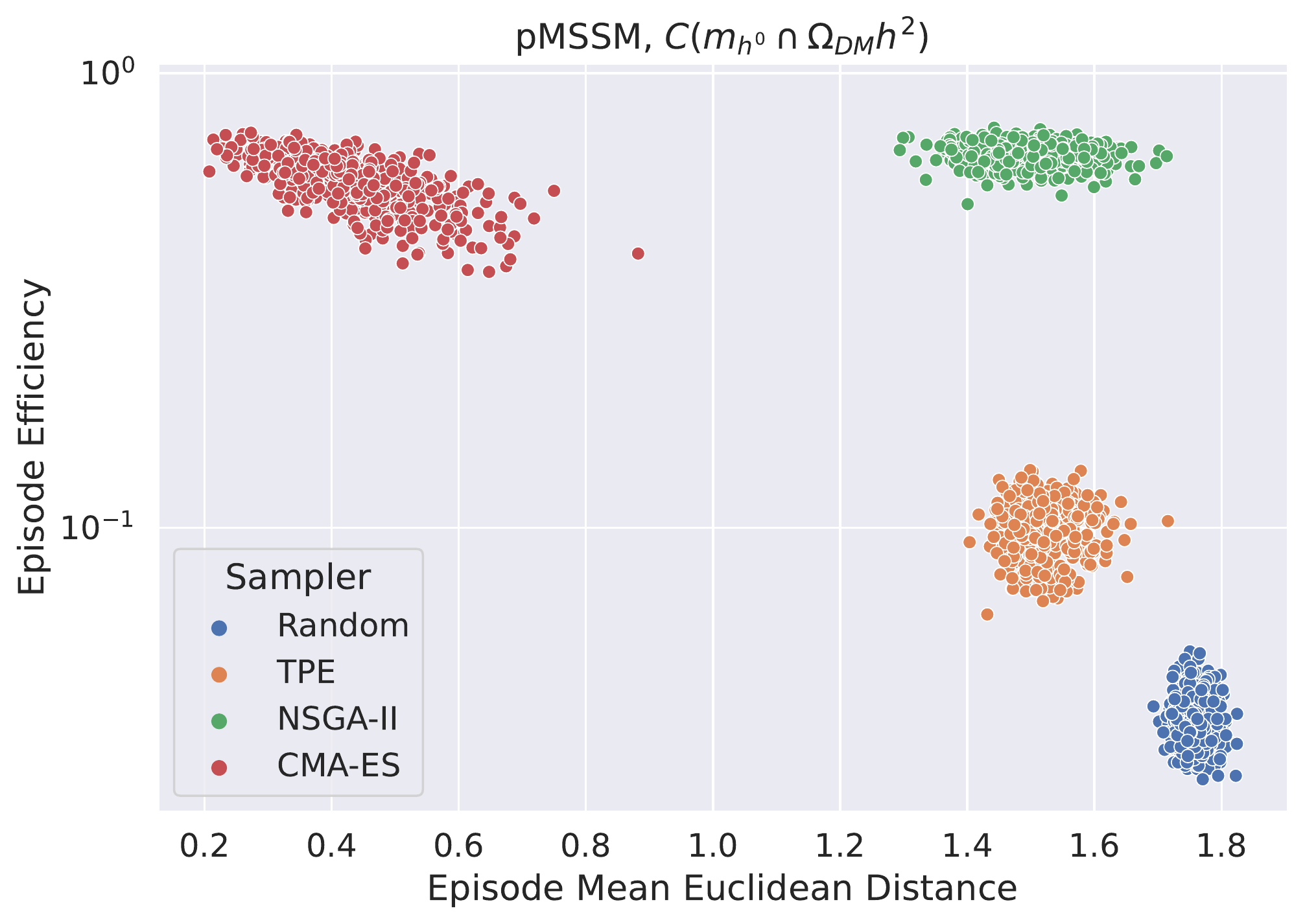}
   \caption{Mean Euclidean distance vs efficiency per episode, for the pMSSM scan with Higgs mass and dark matter relic density constraints.}
  \label{fig:pMSSM_with_mo_euclidean_efficiency}
\end{subfigure}%
\begin{subfigure}{.45\textwidth}
  \centering
  \includegraphics[width=1.0\linewidth]{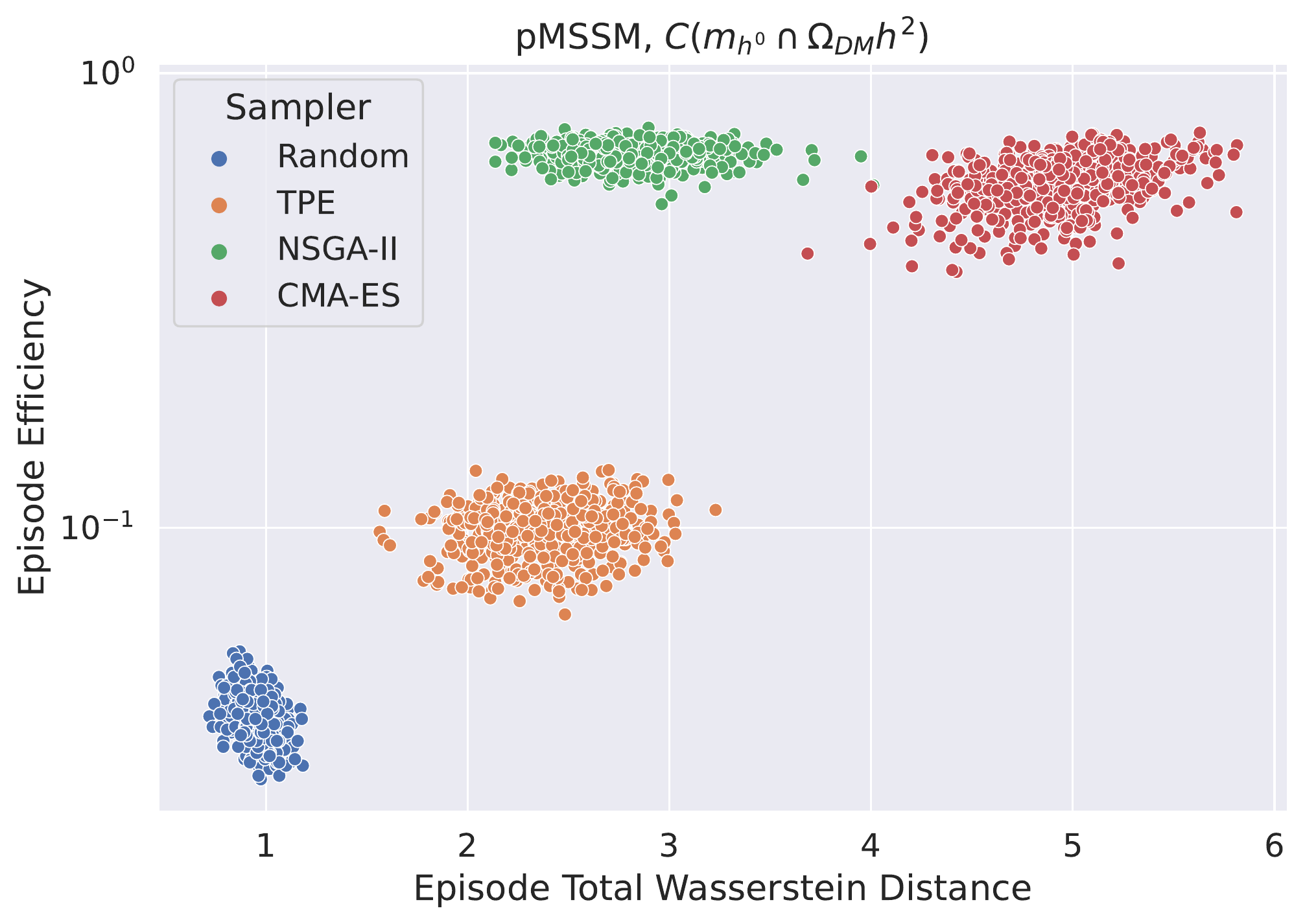}
  \caption{Wasserstein distance vs efficiency per episode, for the pMSSM scan with Higgs mass and dark matter relic density constraints.}
  \label{fig:pMSSM_with_mo_wd_efficiency}
\end{subfigure}
\caption{Efficiency vs Distance metrics, computed using valid points, scatter plots for each sampler for the pMSSM scans.}
\label{fig:pMSSM_scatterplots}
\end{figure}

In~\cref{fig:pMSSM_scatterplots} we present the equivalent plots for the pMSSM scans, where we can observe similar trends and behaviours. Since the pMSSM enjoys greater parametric freedom than the cMSSM, the random sampler has higher sampling efficiency in the case where we consider the dark matter relic density constraint, and there is therefore less room for improvement when comparing to the cMSSM case. However, it is still noticeable that the non-random samplers always improve parameter efficiency, with NSGA-II and CMA-ES already close to the unity efficiency. 

In~\cref{tab:eff,tab:ed,tab:wd} we present the resulting statistics across the different metrics over the episodes. In~\cref{tab:eff} we see that the random sampler has the worst efficiency across all samplers and across all physics cases. For the cases with dark matter relic density constraints, non-random samplers can provide orders of magnitude better sampling efficiency in the best cases, and in the worst case it still more than doubles parameter space efficiency. In general, CMA-ES is the sampler that provides the greatest efficiency, except for the last case where the NSGA-II is the most efficient one. This exception can be explained by the large dimensionality of the pMSSM parameter space combined with the additional constrain of the dark matter relic density, where the CMA-ES sampler struggles to learn the statistics of the valid points due to the \emph{curse of dimensionality}, which plagues shallow Machine Learning components. On the other hand, the NSGA-II sampler does not have any learnt component, making it scale better with the dimension of the parameter space.

\renewcommand{\arraystretch}{1}
\begin{table}
\centering
\begin{tabular}{clllll}
\\ \hline \hline
\multicolumn{1}{l}{}      &                              & \multicolumn{4}{c}{Sampler}                                                      \\  \cline{3-6} 
\multicolumn{1}{l}{Model} & Constraint                   & Random             & TPE                & NSGA-II            & CMA-ES            \\ \hline
\multirow{2}{*}{cMSSM}    & $m_{h}$                      & $0.401 \pm 0.010$  & $0.668 \pm 0.012$ & $0.715 \pm 0.014$  & $\mathbf{0.924 \pm 0.023}$ \\ 
                          & $m_{h_0} \cap \Omega_{DM} h^2$ & $0.006 \pm 0.001$  & $0.127 \pm 0.008$  & $0.281 \pm 0.041$  & $\mathbf{0.687 \pm 0.084}$ \\ \cline{2-6} 
\multirow{2}{*}{pMSSM}    & $m_{h}$                      & $0.309 \pm 0.010 $ & $0.557 \pm 0.038$ & $0.862 \pm 0.015$  & $\mathbf{0.899 \pm 0.034}$ \\ 
                          & $m_{h} \cap \Omega_{DM} h^2$   & $0.038 \pm 0.004$  & $0.099 \pm 0.013$  & $\mathbf{0.663 \pm 0.036}$ & $0.576 \pm 0.073$ \\ \hline \hline
\end{tabular}
\caption{Efficiency statistics for each sampler. The central value and the standard deviation are computed across the episodes. In bold we highlight the best non-random sampler for each physics case.}
\label{tab:eff}
\end{table}

Although efficiency is important, we also want to guarantee that the non-random samplers are properly covering the whole parameter space. In~\cref{tab:ed} we can see the average of the mean euclidean distances. As expected, the random sampler provides the greater mean euclidean distance, meaning that it produces valid points which are quite far apart from each other as a result of the breadth of its sampling. The only exception is for the TPE sampler in the cMSSM without dark matter relic density constraint, this can be due to the Gaussian Mixture Model sampling from two far away centres, even though the result is similar to the random case within the statistical uncertainties. In general we see that the CMA-ES produce points which are very closely together, a result due to its \emph{eager} nature.

\begin{table}
\centering
\begin{tabular}{clllll}
\hline \hline
\multicolumn{1}{l}{}      &                               & \multicolumn{4}{c}{Sampler}                                                                            \\ \cline{3-6} 
\multicolumn{1}{l}{Model} & Constraint                    & Random                   & TPE                     & NSGA-II                 & CMA-ES                  \\ \hline
\multirow{2}{*}{cMSSM}    & $m_{h_0}$                     & $0.686 \pm 0.006$  & $\mathbf{0.706 \pm 0.009}$ & $0.619 \pm 0.017$ & $0.401 \pm 0.058$ \\
                          & $m_{h_0} \cap \Omega_{DM}h^2$ & $0.625 \pm 0.075$  & $\mathbf{0.376 \pm 0.032}$ & $0.321 \pm 0.066$ & $0.223 \pm 0.097$ \\ \hline
\multirow{2}{*}{pMSSM}    & $m_{h_0}$                     & $1.745 \pm 0.006 $ & $\mathbf{1.659 \pm 0.021}$ & $1.594 \pm 0.048$ & $0.750 \pm 0.128$ \\
                          & $m_{h_0} \cap \Omega_{DM}h^2$ & $1.758 \pm 0.021$  & $\mathbf{1.523 \pm 0.041}$ & $1.500 \pm 0.069$ & $0.437 \pm 0.097$ \\ \hline \hline
\end{tabular}
\caption{Mean Euclidean Distance of valid points statistics for each sampler. The central value and the standard deviation are computed across the episodes. In bold we highlight the best non-random sampler per physics case.}
\label{tab:ed}
\end{table}

Regarding the Wasserstein distance statistics in~\cref{tab:wd}, we observe similar trends. I.e., the random sampler is the sampler that provides the widest coverage of the parameter space as it is the one producing parameter distributions closer to a uniform distribution. The sampler that produces the most distorted distributions is CMA-ES, a phenomenon linked to its cluster points highlighted in the table above, with TPE providing on average the best coverage out of the non-random samplers. NSGA-II appears just behind TPE, namely it provides similar results to TPE in the cases where the dark matter relic density is switched on.

\begin{table}
\centering
\begin{tabular}{clllll}
\hline\hline
\multicolumn{1}{l}{}      &                               & \multicolumn{4}{c}{Sampler}                                                    \\ \cline{3-6} 
\multicolumn{1}{l}{Model} & Constraint                    & Random             & TPE               & NSGA-II           & CMA-ES            \\ \hline
\multirow{2}{*}{cMSSM}    & $m_{h_0}$                     & $0.229 \pm 0.014$  & $\mathbf{0.276 \pm 0.020}$ & $0.364 \pm 0.039$ & $0.627 \pm 0.099$ \\
                          & $m_{h_0} \cap \Omega_{DM}h^2$ & $0.797 \pm 0.113$  & $\mathbf{1.101 \pm 0.062}$ & $1.150 \pm 0.090$ & $1.186 \pm 0.117$ \\ \cline{2-6} 
\multirow{2}{*}{pMSSM}    & $m_{h_0}$                     & $0.495 \pm 0.030 $ & $\mathbf{1.270 \pm 0.137}$ & $2.028 \pm 0.188$ & $3.997 \pm 0.449$ \\
                          & $m_{h_0} \cap \Omega_{DM}h^2$ & $0.939 \pm 0.079$  & $\mathbf{2.369 \pm 0.274}$ & $2.800 \pm 0.278$ & $4.932 \pm 0.331$ \\ \hline \hline
\end{tabular}
\caption{Wasserstein Distance computed on valid points statistics for each sampler. The central value and the standard deviation are computed across the episodes. In bold we highlight the best non-random sampler per physics case.}
\label{tab:wd}
\end{table}

\begin{figure}
\centering
\begin{subfigure}{.475\textwidth}
  \centering
  \includegraphics[width=1.0\linewidth]{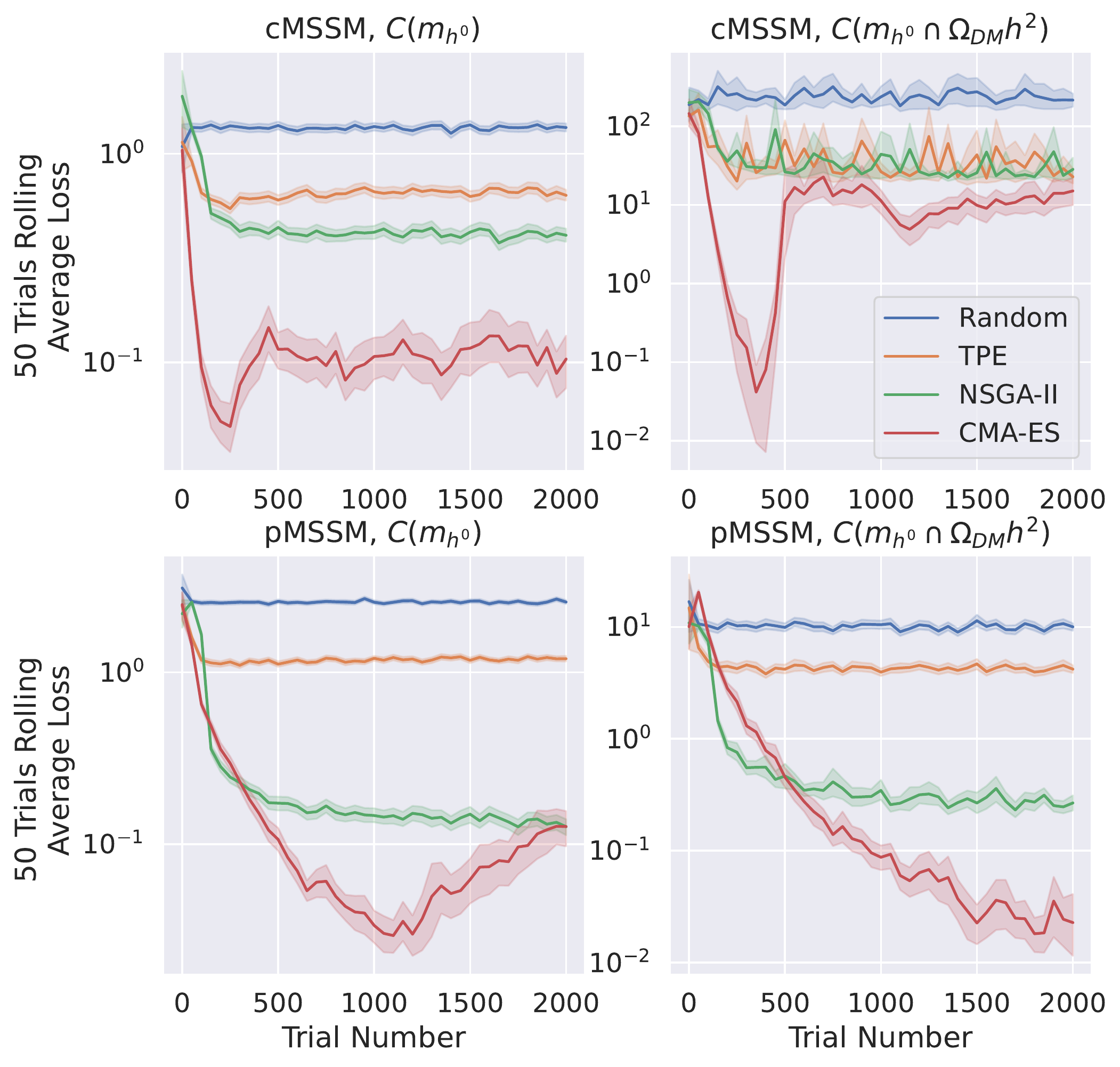}
   \caption{Rolling average loss per trial.}
  \label{fig:rolling_value}
\end{subfigure}%
\begin{subfigure}{.475\textwidth}
  \centering
  \includegraphics[width=1.0\linewidth]{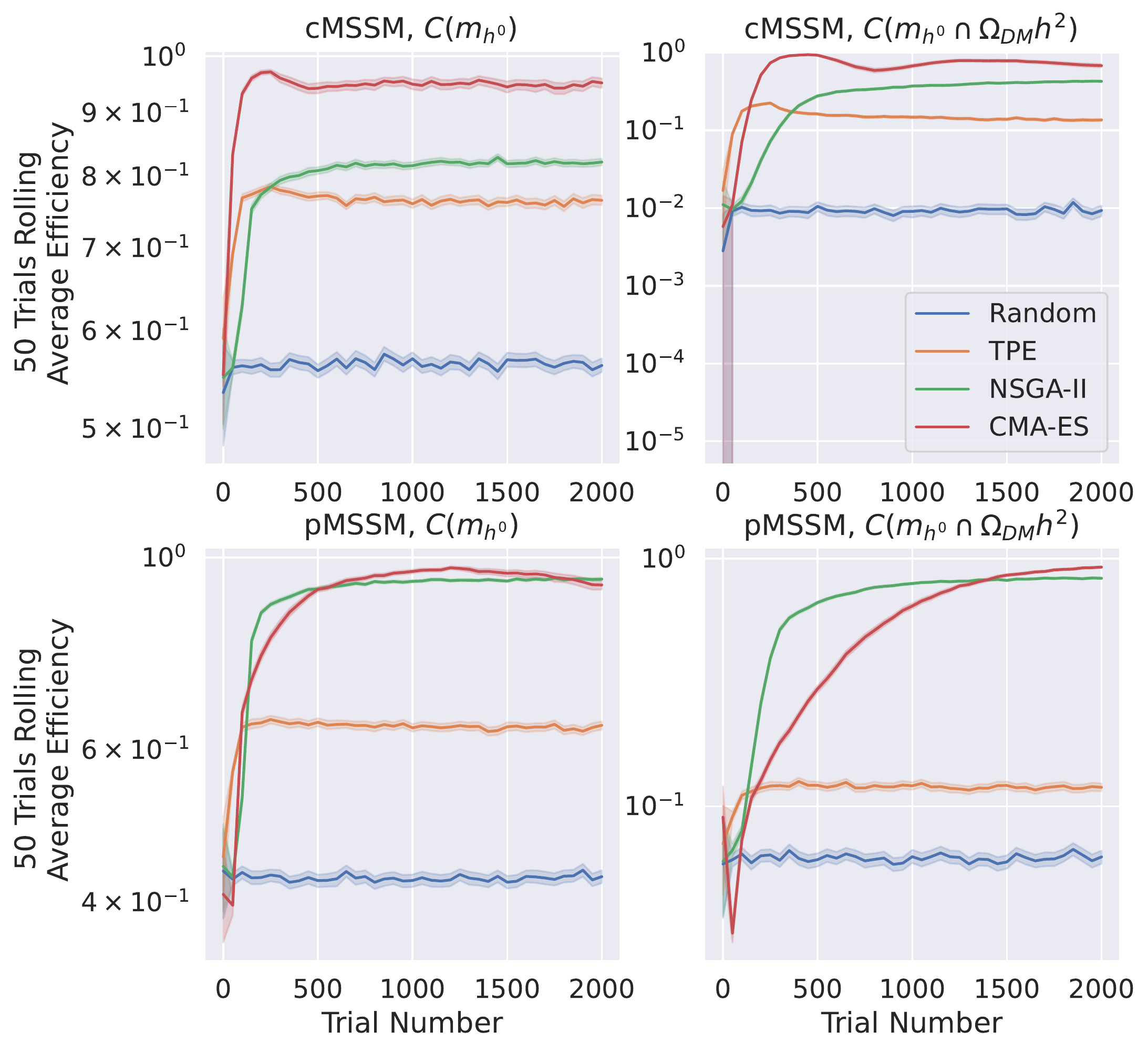}
  \caption{Rolling average efficiency per trial.}
  \label{fig:rolling_efficiency}
\end{subfigure}
\caption{Rolling metrics history for each sampler. Each metric is computed in each episode as a function of the previous 50 trials, and the shaded bands represent 95\% confidence intervals computed over the 500 episodes.}
\label{fig:rolling_value_and_efficiency}
\end{figure}

Another important aspect to compare different samplers is to see how fast they converge to valid regions, as the non-random samplers work sequentially, improving the quality of a suggested point with respect to the points it has suggested before. In order to assess this, we present in~\cref{fig:rolling_value_and_efficiency} the rolling average values for the loss, c.f.~\cref{eq:loss}, and the efficiency as a function of the number of trials.

In~\cref{fig:rolling_value} we see that the random sampler average loss value is constant over time. This is expected, as each sampled point of the random sampler is independent of any other sampled point. The same is not the case for the non-random samplers, as they attempt to produce ever better points that minimise the loss. This is explicitly observable in these plots, as we see the average loss decreasing considerably after just a few trials. Indeed, for most cases the average loss stabilises just after a few trials, and always below the average loss of the random sampler, showing how these samplers keep producing points which are on average better than those sampled by the random sampler. The CMA-ES presents the most different behaviour, with a rapid dip followed by an increase of the average loss in all cases except for the pMSSM with the dark matter relic density constraint. This behaviour is understood as we switch on the \verb|restart_strategy| flag, which will restart the population once it is seemingly in a local minima in order to increase exploration. For the pMSSM with the dark matter relic density constraint case, we observe that the CMA-ES does not seem to converge within the 2000 trials allowance, and keeps suggesting better solutions. This can be due to the fact that CMA-ES works with a multivariate normal from which it samples points in this highly dimensional space, and therefore is challenged by the \emph{curse of dimensionality} as this might not be the most appropriate learnable model for such a high dimensional space.

In the plot for the rolling efficiency,~\cref{fig:rolling_efficiency}, we observe a complementary behaviour. All non-random sampler quickly saturate their sampling efficiency in almost all the cases. The exceptions are once again related to CMA-ES. For all the cases except for the pMSSM with the dark matter relic density constraint, the CMA-ES restarts its sampling after hitting an optimal sampling efficiency. For the other case, it has yet to achieve that optimal sampling efficiency point within the 2000 trials allowance within each episode.

It is interesting to point out how narrow the 95\% confidence intervals are. Meaning that for sampler, each episode has a similar evolution, allowing to draw the conclusions above.

\begin{figure}
\centering
\begin{subfigure}{.45\textwidth}
    \centering
    \includegraphics[width=.8\linewidth]{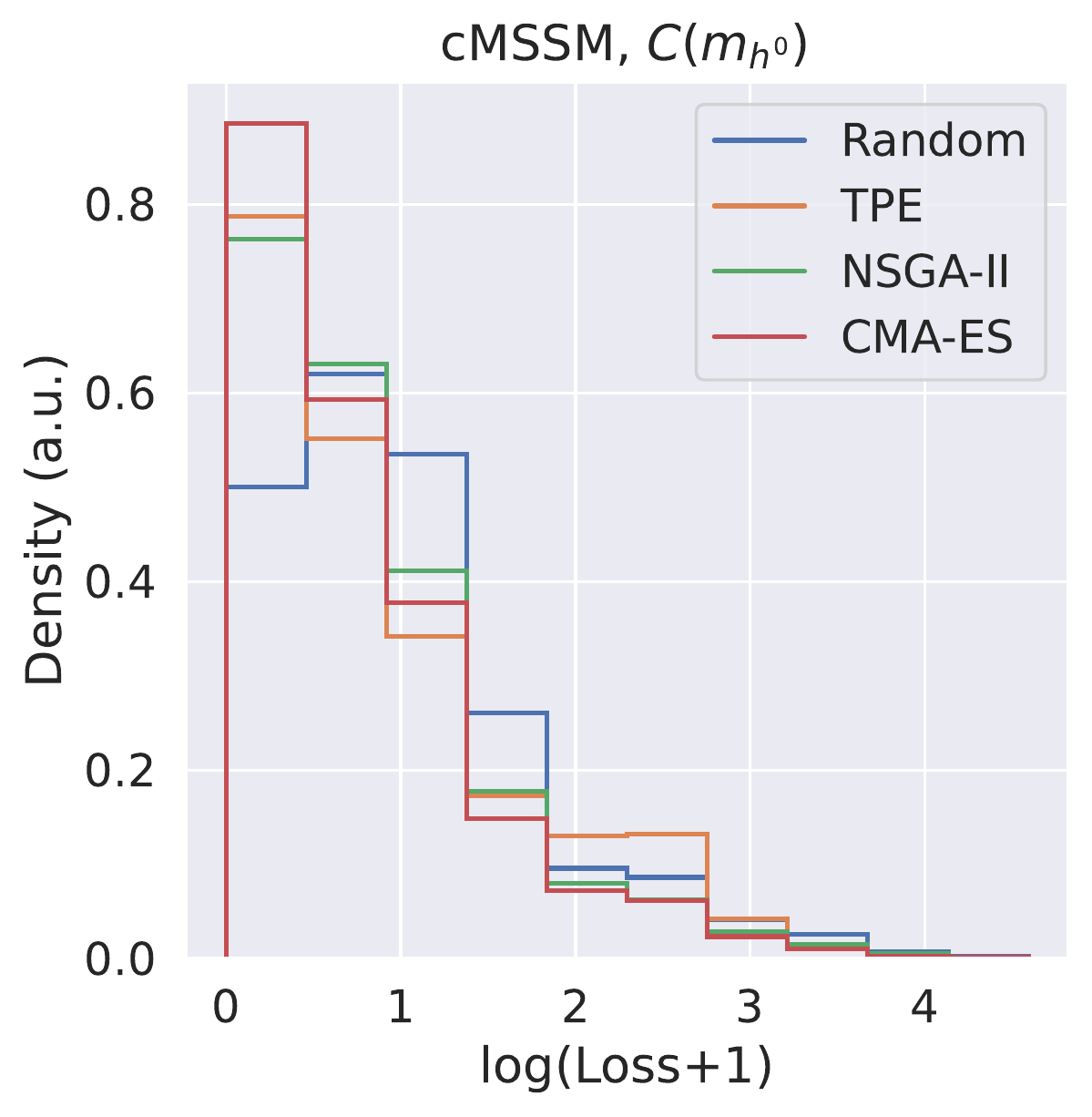}
\end{subfigure}
\begin{subfigure}{.45\textwidth}
    \centering
    \includegraphics[width=.8\linewidth]{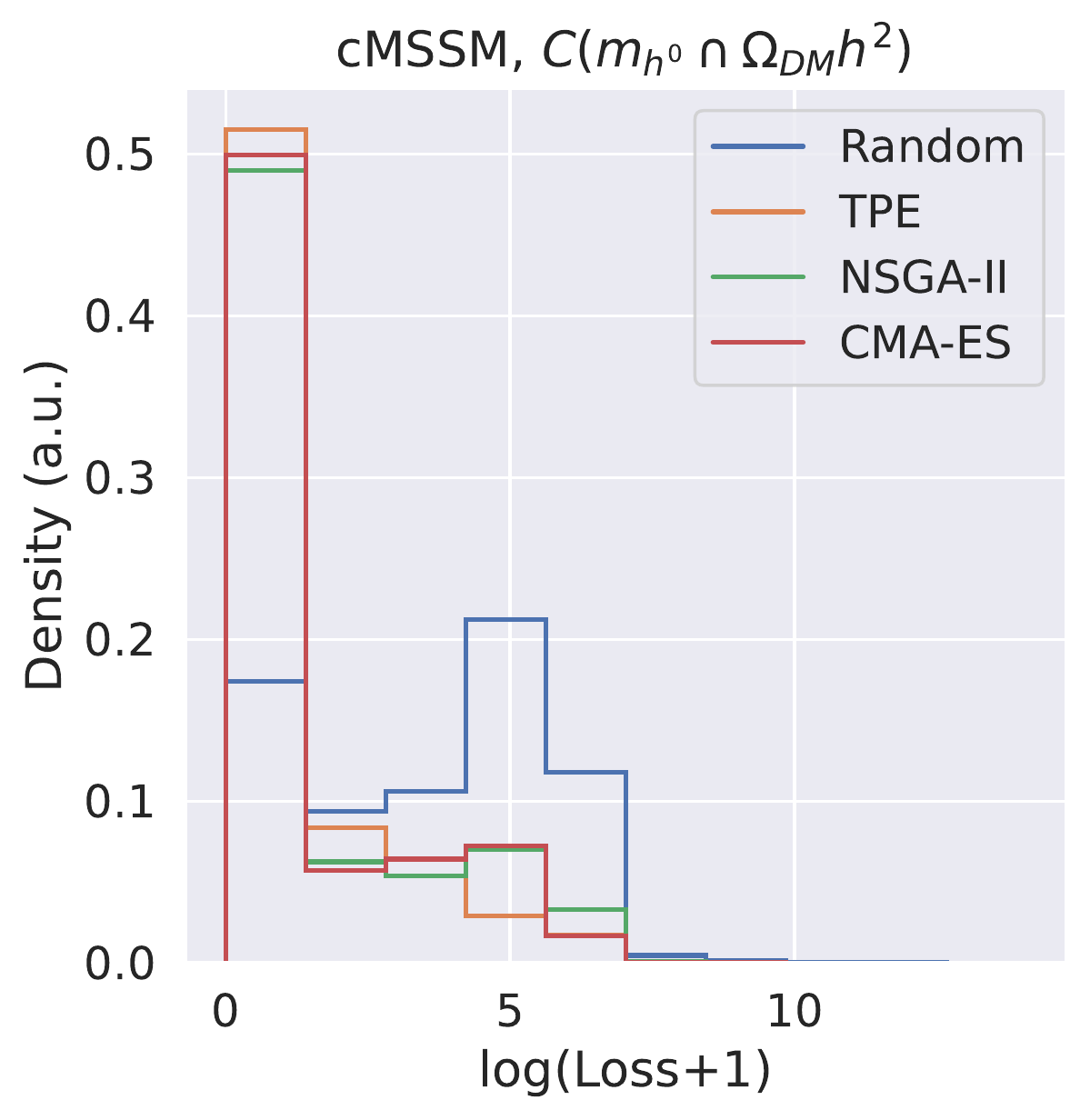}
\end{subfigure}\\
\begin{subfigure}{.45\textwidth}
    \centering
    \includegraphics[width=.8\linewidth]{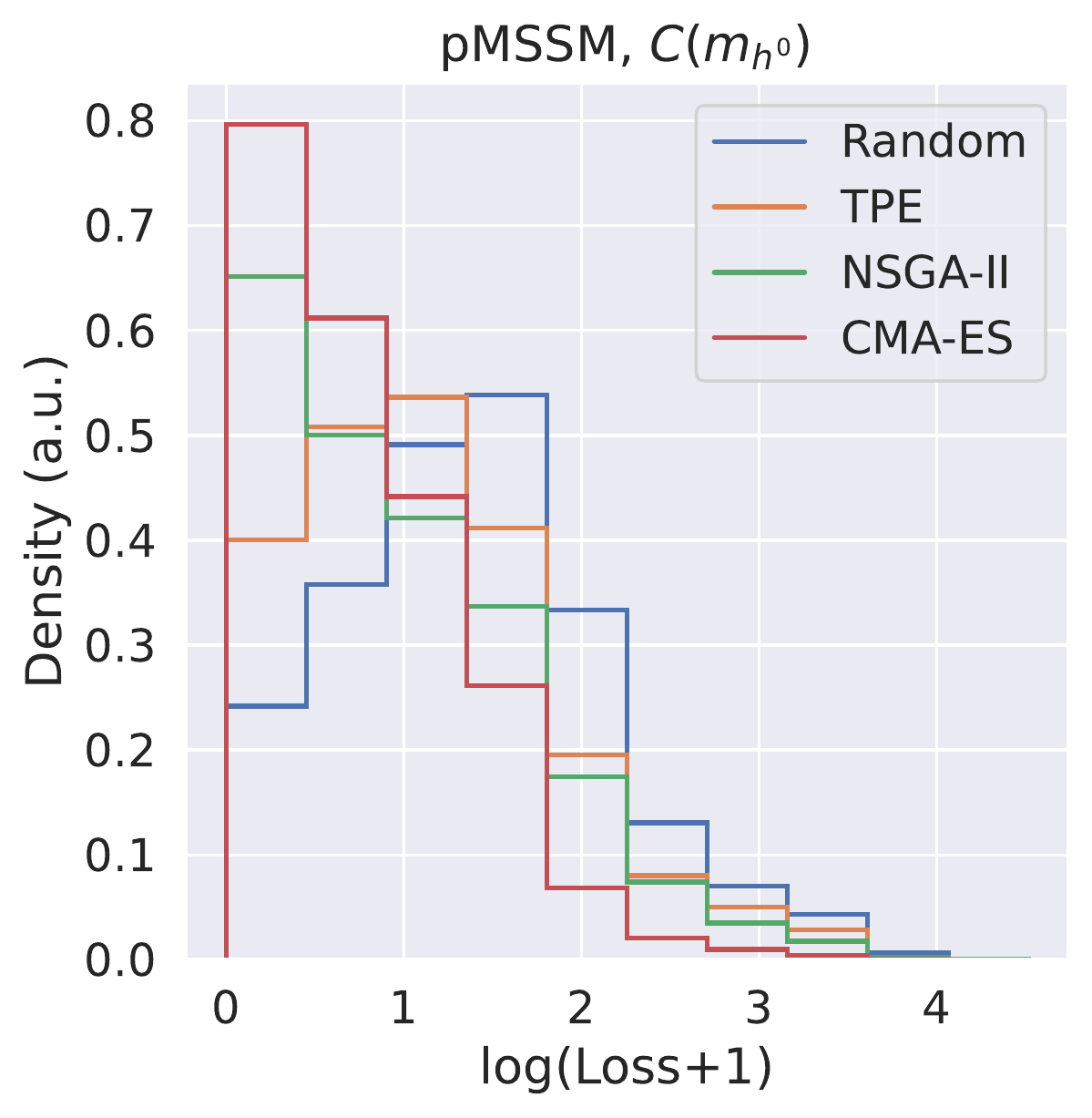}
\end{subfigure}
\begin{subfigure}{.45\textwidth}
    \centering
    \includegraphics[width=.8\linewidth]{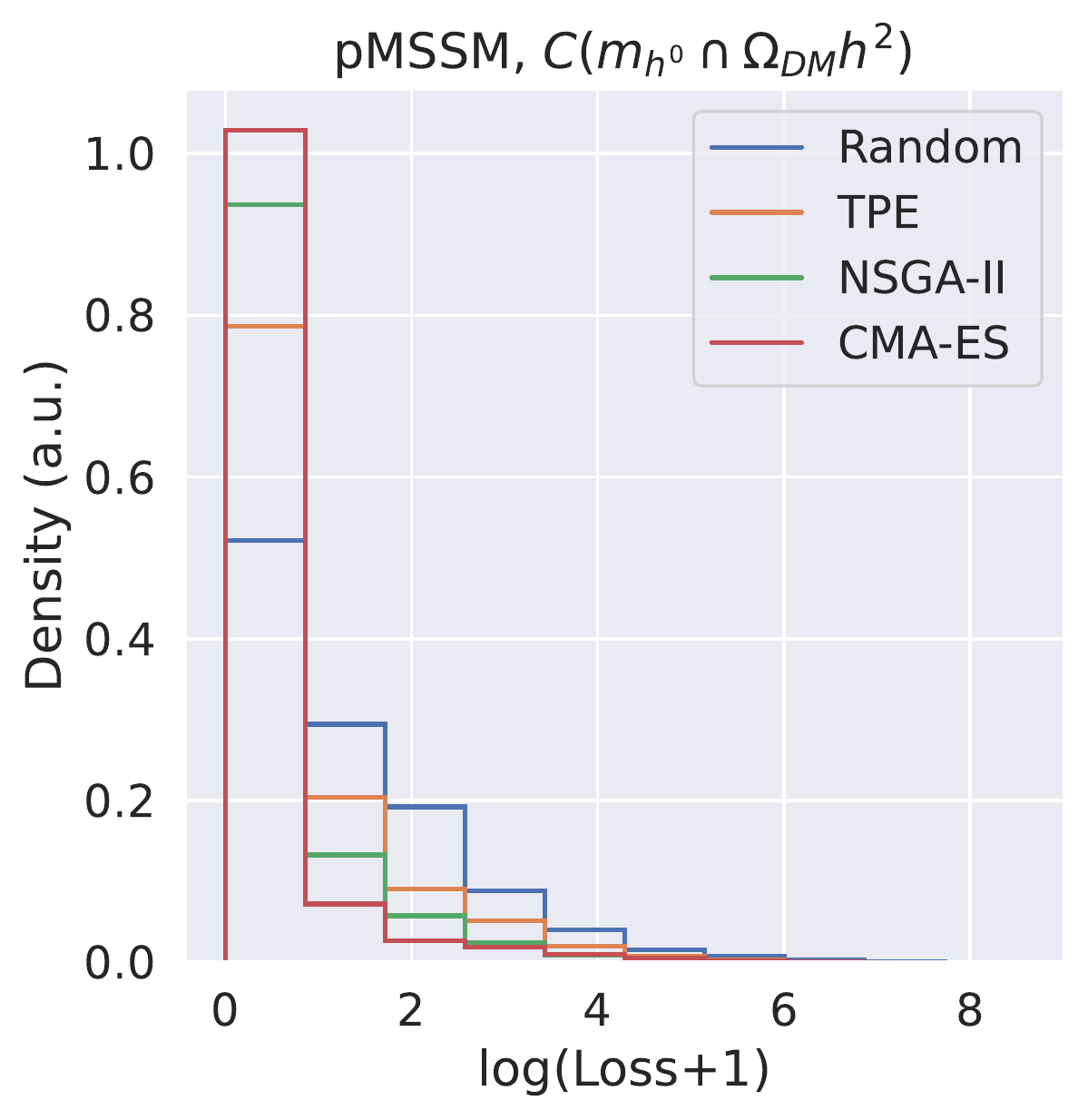}
\end{subfigure}
\caption{Distributions of loss values for non-valid, but physical, points for each sampler and for the different physiscs cases.}
\label{fig:loss_non_valid_points}
\end{figure}

The above trial evolution plots show that the samplers progressively improve the quality of the suggested points, as measure by how likely they are to minimise the loss function. This also suggests that points that have not satisfied the conditions, but are otherwise physical (i.e. that they have successfully produced a spectrum and a dark matter candidate), should have lower loss values than points randomly sampled. In~\cref{fig:loss_non_valid_points} we see the distribution of the values of the loss function for non-valid, albeit physical, points. We see that for all the physics cases, the values of the losses are always lower for non-random samplers than for the random sampler. This is in agreement with the expectation that non-valid points suggested by the non-random samplers are closer to be valid than those sampled from a random sampler.

\subsection{Sampling Time}

\begin{figure}
    \centering
    \includegraphics[width=0.8\linewidth]{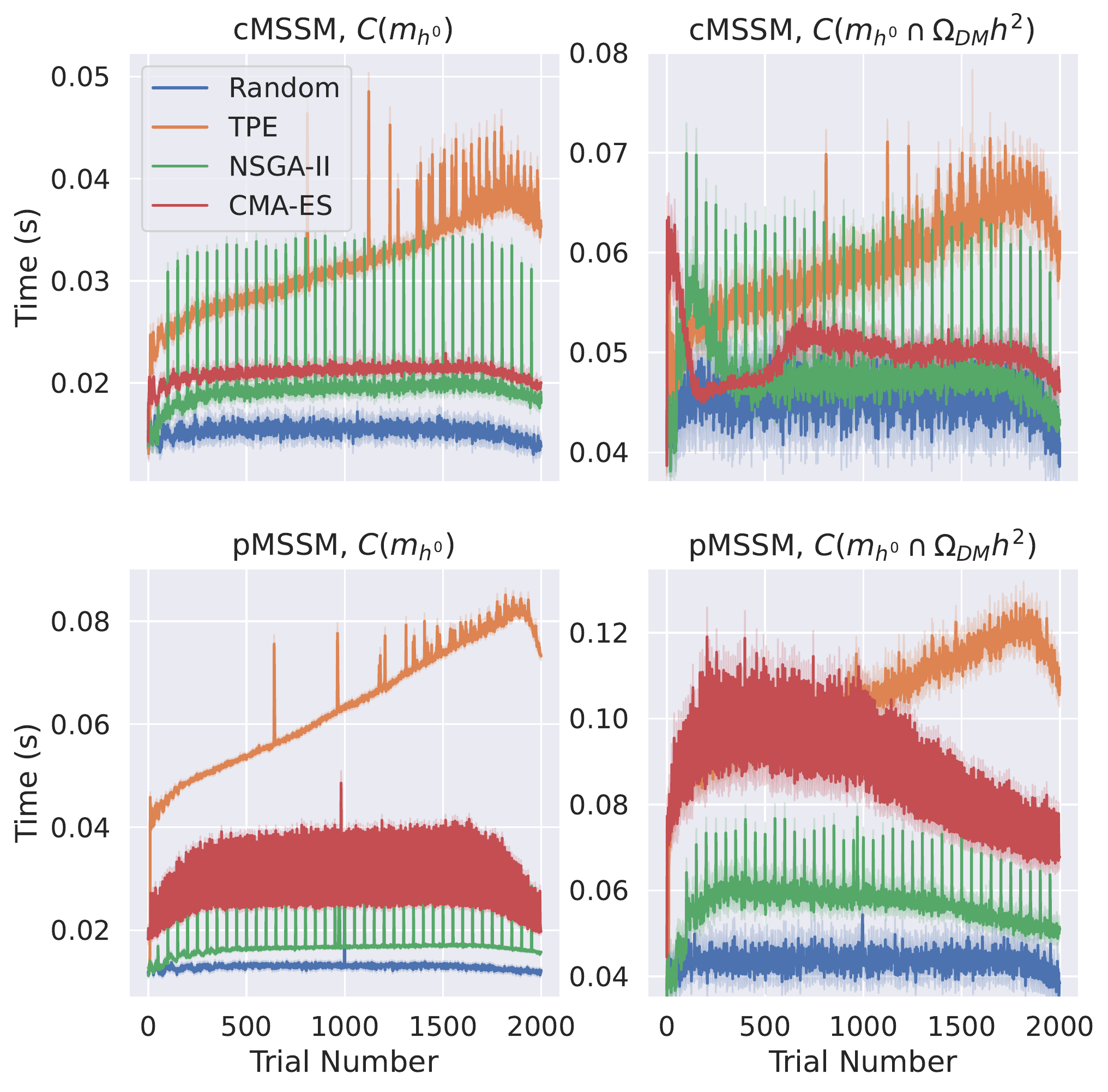}
    \caption{Trial evolution time over the episode for each sampling algorithm. The shaded region represents 95\% confidence intervals.}
    \label{fig:trial_time}
\end{figure}

We have already shown that the non-random samplers drastically improve sampling efficiency over the random sampler. However, the methodology and algorithms presented in this work are only useful if the non-random samplers do not impose a computational overhead that would make these scans impractically slow. In~\cref{fig:trial_time} we show the trial evolution time over the episodes. These plots present an artificial deformation that does not originate from our methodology: the reduction of trial time at the end of the episodes. This is due to the fact that various episodes were executed in parallel, leading to concurrency competition when reading and writing to the hard-drive, and as episodes finished it became faster to complete those still running.

In all physics cases, the random sampler is the fastest, which is expected as it does not include any new sampling algorithm. For all non-random sampler cases, we observe an increase of per-trial evaluation time due to the added computational overhead of the algorithm.

For all physics cases, we witness the linear growth in time for the TPE, which is in line with our expectations as the TPE fits a Gaussian Mixture Model that has a computational complexity that grows linearly with the number of points. This also means that the total running time of an episode, being the sum of all trials time in that episode, grows quadratically with the maximum number of trials in the episode. This is the reason why we restricted to a maximum of 2000 trials per episode, as this quadratic run-time growth, which prevents very long episodes, was identified early on in our study. This also means that for a specific problem where TPE cannot find valid points within the first few thousands trials, it will likely not be a good sampler to perform a thorough scan as its run-time will become prohibitively slow.

An interesting observation regards the spikes in time of the NSGA-II every 50 trials, giving it a comb-like shape. This happens as the default population size is 50, for which after 50 trials the algorithm has to perform the genetic operations over such trials -- sorting, selection, cross-over, and mutation -- in order to produce candidate points to be evaluated in the following 50 trials. Despite these spikes, the NSGA-II presents the lightest overhead, being constantly the fastest sampler after the random sampler.

\section{Conclusions\label{sec:conclusions}}

In this work we have reframed the parameter space scanning task for validation of BSM models as a black-box optimisation problem. To accomplish this, we retain the information of an invalid point and how \emph{far} it is from being valid using a loss function that can then be minimised using black-box optimisation algorithms from the Artificial Intelligence and Machine Learning literature. We introduced three of such algorithms: Tree-Parzen Estimator, a Bayesian optimisation algorithm; Nondominated Sorting Genetic Algorithm II, a genetic algorithm; and the Covariance Matrix Adaptation Evolution Strategy, a non-genetic evolutionary algorithm. These algorithms search for valid points by interacting with the loss function, which in turn is computed using the produced observables obtained from the computational routines. In this work, we focused on the physics cases of the cMSSM and the pMSSM, with and without the further constraint of having a valid candidate for dark matter.

The novel approach presented tackles the shortcomings of current methodologies which rely on a vast collection of valid points before they can be used to sample new points, which can be a challenge for scanning tasks where random sampling can be highly inefficient from the start. Furthermore, by not being equivalent to a fit to likelihoods, our approach can be used with bounds that are derived from theory as well as experimental limits on new physics, which are two common constraints used in BSM constraining scans that do not have a corresponding likelihood. 

We showed that this approach, not requiring any a priori knowledge of the parameter space, provides orders of magnitude better sampling efficiencies in comparison with the random sampling strategy usually employed for this task. We showed that this benefit comes at a trade-off cost between efficiency and coverage of the parameter space, with different samplers providing distinct realisations of this trade-off: the TPE provides results similar to the random sampler, while the CMA-ES can achieve near-unity sampling efficient, and finaly the NSGA-II finds its place somewhere in-between these two in terms of \emph{exploration-exploitation} trade-off. Consequently, the best sampler will greatly depend on the task at hand and how difficult it is, as well as the goals of the BSM model builder in a specific study. For example, if the scan is performed on highly dimensional parameter spaces the evolutionary algorithm, NSGA-II, is better suited since it does not suffer from the \emph{curse of dimensionality} while providing a middle ground between \emph{exploration} and \emph{exploitation}; if the problem revolves around a highly constrained model, where the random sampler has little efficiency, in a small dimension parameter space, then the CMA-ES would be a better choice, as it converges quickly to valid regions of the parameter space do to its \emph{eager} nature; finally, the Bayesian algorithm, TPE, provides results more similar to the random sampler, and should therefore preferred when coverage is the main concern, although it will struggle to find good points if it fails to converge to a valid region within the first few thousand points due to its run-time becoming prohibitively slow.

Although we have shown the great potential benefit of using non-random samplers to perform parameter space scanning of BSM models, our work also points at future directions to improve upon the proposed methodologies. First, despite choosing some options that differ from the default parameters, we have not undertaken any optimisation of the samplers, which could further improve the presented metrics. Secondly, we have to reiterate that the proposed algorithms were not designed for the specific case of BSM parameter space scan and constraining -- which requires extensive coverage over highly multidimensional spaces --, and therefore there is the potential to further improve them, or design new ones, that can mitigate the \emph{exploration-exploitation} cost of choosing one side over the other, or the sensitivity to the \emph{curse of dimensionality} of same of the samplers. Finally, we made an explicit choice of summing together two constraint functions instead of optimising each separately as a \emph{multi-objective} optimisation problem. This choice was made so that we could use different optimisers that cannot perform such task, such as the CMA-ES, but it is likely that algorithms like NSGA-II, which were designed especially for such problems, will provide even better samplers for problems that involve multiple joint constraints.

Finally, we notice that the methodology herein is not restricted to SUSY model building, and can be used with any computational routine and set of constraints -- regardless the BSM framework and computational language where the routines are written -- and therefore provides a general new paradigm for parameter space scanning and BSM model validation.

\section*{Acknowledgements}
We thank José Santiago Pérez and Jorge Romão for the careful reading of the paper draft and for the useful discussions.
This work is supported by FCT - Fundação para a Ciência e a Tecnologia, I.P. under project CERN/FIS-PAR/0024/2019. FAS is also supported by FCT under the research grant with reference UI/BD/153105/2022. The computational work was partially done using the resources made available by RNCA and INCD under project CPCA/A1/401197/2021.
\appendix

\bibliography{references}{}
\bibliographystyle{unsrt}

\end{document}